\documentclass[aps,pra,reprint,superscriptaddress]{revtex4-2}

\usepackage{amsmath}
\usepackage{graphicx}
\usepackage{esint}
\usepackage[colorlinks=true,linkcolor=blue,citecolor=blue,urlcolor=blue]{hyperref}
\usepackage{orcidlink}
\usepackage{tikz}

\let\origorcidlink\orcidlink
\renewcommand\orcidlink[1]{%
	\raisebox{0.32\height}{\scalebox{0.75}{\origorcidlink{#1}}}%
}

\begin{document}

\title{Numerical evaluation of Casimir forces \\using the discontinuous Galerkin time-domain method}

\author{Carles~Mart\'i~Farr\`{a}s\,\orcidlink{0009-0001-2764-7205}}
\email[]{marticar@physik.hu-berlin.de}
\affiliation{Institut~f{\"u}r~Physik,~Humboldt Universit{\"a}t~zu~Berlin,~Newtonstra\ss{}e~15,~12489~Berlin,~Germany}

\author{Bettina~Beverungen\,\orcidlink{0000-0002-6701-4269}}
\affiliation{Institut~f{\"u}r~Physik,~Humboldt Universit{\"a}t~zu~Berlin,~Newtonstra\ss{}e~15,~12489~Berlin,~Germany}

\author{Philip~Tr\o{}st~Kristensen\,\orcidlink{0000-0001-5804-1989}}
\affiliation{DTU Electro,~Technical~University~of~Denmark,~\O{}rsteds~Plads~343,~2800~Kgs.~Lyngby,~Denmark}

\author{Francesco~Intravaia\,\orcidlink{0000-0001-7993-4698}}
\affiliation{Institut~f{\"u}r~Physik,~Humboldt Universit{\"a}t~zu~Berlin,~Newtonstra\ss{}e~15,~12489~Berlin,~Germany}

\author{Kurt~Busch\,\orcidlink{0000-0003-0076-8522}}
\affiliation{Institut~f{\"u}r~Physik,~Humboldt Universit{\"a}t~zu~Berlin,~Newtonstra\ss{}e~15,~12489~Berlin,~Germany}
\affiliation{Max-Born-Institut,~Max-Born-Stra\ss e~2a,~12489~Berlin,~Germany}


\begin{abstract}
	We present a time-domain scheme for computing Casimir forces within the Maxwell stress tensor formalism, together with a specific realization using the finite-element-based discontinuous Galerkin time-domain method. The approach enables accurate evaluation of Casimir--Lifshitz interactions for a wide range of geometries and material properties at finite temperature. At the core of the method, the electromagnetic Green's tensor is expressed as the system's response to dipolar excitations, thereby recasting the Maxwell stress tensor into a set of classical scattering problems driven by electric and magnetic dipoles. We validate the approach against reference calculations of the Casimir interaction between parallel half-spaces at both zero and nonzero temperature. We further demonstrate its applicability to finite, cylindrically symmetric geometries for which closed-form solutions are unavailable, obtaining accurate agreement with asymptotic predictions based on physical considerations. These findings illustrate the method's potential for studying Casimir interactions in realistic micro- and nanoscale structures, relevant to nanodevice design and experimental settings.
\end{abstract}

\maketitle

\section{Introduction\label{sec:intro}}

Fluctuation-induced phenomena, arising from quantum and thermal fluctuations of the electromagnetic field, lead to fascinating effects that extend beyond the domain of classical electrodynamics. A prominent example is the Casimir effect, a typically attractive force between electrically neutral macroscopic objects. First predicted by H. B. G. Casimir for perfectly reflecting mirrors in vacuum and attributed to zero-point fluctuations of the quantum electromagnetic field~\cite{Casimir1948a}, it was later extended by Lifshitz to realistic materials, incorporating properties such as dispersion and dissipation at finite temperature~\cite{Lifshitz1956}. The Casimir force can be regarded as a macroscopic manifestation of the entire system's quantum fluctuations, including those within the materials. In this sense, it encompasses van der Waals interactions, originally introduced by van der Waals and later described by London as an attractive force between closely spaced particles with fluctuating dipole moments~\cite{London2000}. Casimir and Polder extended London's theory to include retardation effects, describing the interaction between an atom and a macroscopic surface~\cite{Casimir1948}, with the resulting fluctuation-induced phenomenon commonly referred to as the Casimir--Polder interaction.

While negligible at macroscopic scales, Casimir forces become significant at submicrometer distances, rapidly increasing as the separation decreases. Beyond its theoretical relevance, the Casimir effect has been consistently confirmed through experimental observations~\cite{Blokland1978}. Modern measurements rely on precision instruments such as torsion pendulums~\cite{Lamoreaux1997}, atomic force microscopes~\cite{Mohideen1998,Roy1999,Harris2000}, and microelectromechanical torsional oscillators~\cite{Decca2003,Decca2005,Decca2007,Krause2007}. These setups frequently employ curved surfaces in close proximity to a substrate in order to minimize alignment errors. The interaction is then estimated using the proximity-force approximation (PFA)~\cite{Derjaguin1934}, which describes it as a sum of infinitesimal parallel-plate elements. Successful plane--plane measurements have also been reported~\cite{Bressi2002}. Advances in experimental sensitivity have brought the detection of finite-temperature corrections within reach~\cite{Sushkov2011,Bezerra2006,Bimonte2021,Bimonte2016,Haghmoradi2024}, although some related controversies remain unresolved~\cite{Decca2005,Klimchitskaya2011,Banishev2013}. From a technological standpoint, the ongoing miniaturization of nanostructured devices has made the Casimir effect increasingly relevant in the design of nano- and micro-electromechanical systems, sparking interest within both the experimental and theoretical research communities~\cite{Gong2021,Shen2025}. One consequence is that, due to its typically attractive nature, the Casimir effect can lead to stiction --- undesired adhesion of movable components --- thereby disrupting device performance~\cite{Serry1998,Chan2001,Buks2001}.

\begin{figure}[b]
	\includegraphics[width=\linewidth]{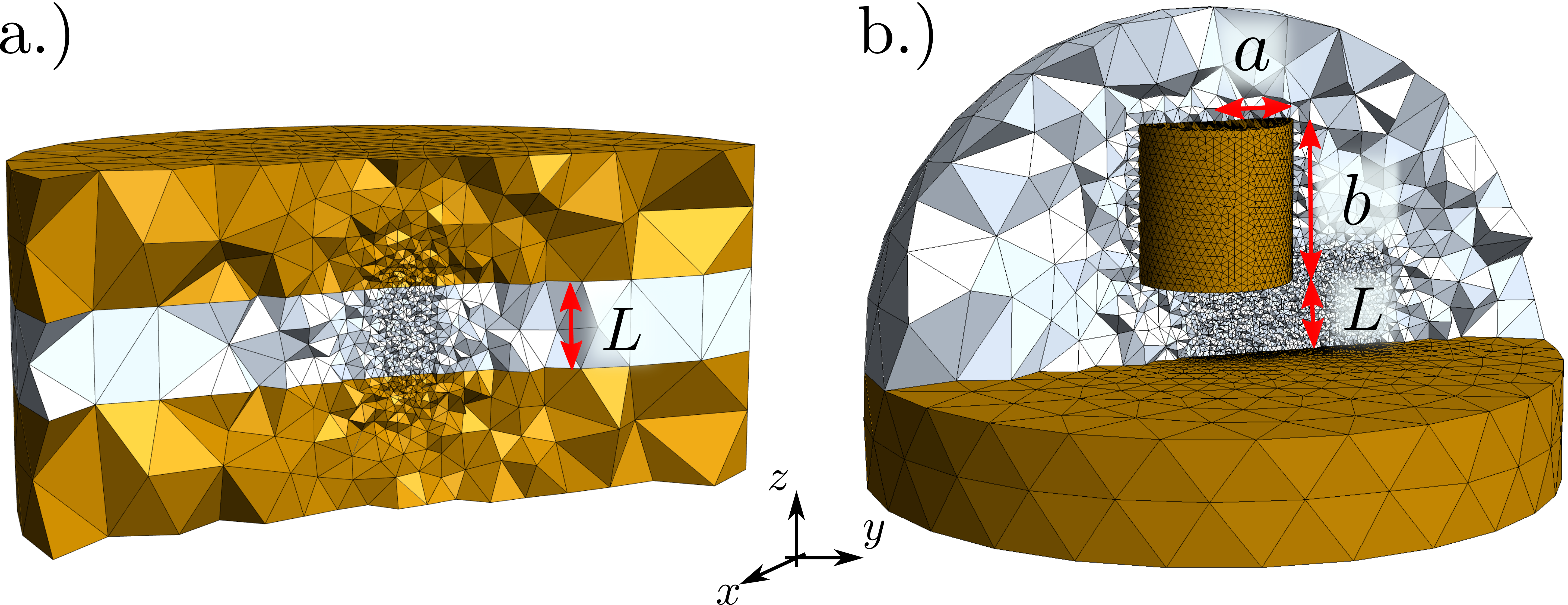}
	\caption{Representative tetrahedral meshes generated with Gmsh~\cite{Geuzaine2009} for (a) plane--plane and (b) cylinder--plane geometries. $L$ is the minimum surface-to-surface separation, while $a$ and $b$ denote the cylinder radius and height, respectively.\label{fig:2hs_hs-cyl_mesh}}
\end{figure}

In this context, theoretical descriptions of Casimir interactions are essential for understanding geometry- and material-dependent effects, designing nanostructured devices, and optimizing experimental setups. Over time, semi-analytical approaches based on scattering methods have been developed to describe practical systems, including complex geometries, realistic material models, and finite-temperature corrections~\cite{Kenneth2008,Rahi2009,Maghrebi2011,Hartmann2017,Zandi2010,CanaguierDurand2010,CanaguierDurand2010a}. However, these methods remain strongly geometry-dependent. As an alternative, numerical techniques provide a more flexible framework for treating arbitrary three-dimensional configurations without increasing the mathematical complexity. Common approaches include boundary-element methods~\cite{Xiong2009,Xiong2010,Reid2011,Reid2013}, as well as finite-difference techniques in both the frequency domain (FDFD)~\cite{Rodriguez2007} and the time domain (FDTD)~\cite{Rodriguez2009,McCauley2010}. These approaches incorporate temperature effects~\cite{Pan2011}, a broad range of material models, and arbitrary geometries, revealing interesting phenomena such as nonmonotonic force behavior~\cite{Rodriguez2007a,Rahi2008}, geometry-induced repulsion for cylindrical objects above a perforated plate~\cite{Levin2010,McCauley2011,Reid2013}, or material-induced suspension in fluids~\cite{Rodriguez2008a,Rodriguez2013}. In this context, achieving numerically accurate Casimir force calculations for fully three-dimensional systems with arbitrary geometries and material properties, while consistently incorporating finite-temperature effects, remains technically challenging due to the presence of multiple relevant length scales.

In this work, we extend the numerical framework introduced by Kristensen et al.~\cite{Kristensen2023} for high-accuracy Casimir--Polder calculations to include both Casimir interactions and finite-temperature corrections within the Maxwell stress tensor formalism. Our approach employs the discontinuous Galerkin time-domain (DGTD) method~\cite{Busch2011}, a finite-element-based technique for numerically solving Maxwell's equations in the time domain. As representative examples, we consider the geometries shown in Fig.~\ref{fig:2hs_hs-cyl_mesh}. First, we examine the Casimir interaction between two metallic half-spaces and compare our results with established semi-analytical solutions~\cite{Lifshitz1956}. We further consider the interaction between a finite cylindrical object and a half-space, a configuration previously addressed semi-analytically only for a perfectly conducting disk with a radius smaller than the separation distance~\cite{Emig2016}. The finite cylinder lacks the symmetry necessary for an efficient semi-analytical treatment. Although rapidly converging multiple-scattering expansions have recently been reported~\cite{Bimonte2023,Emig2023}, a semi-analytical solution for a dispersive and dissipative finite cylinder remains unavailable. Consequently, we validate our results by confirming the expected asymptotic behavior at both large and small separations relative to the cylinder radius.

The present framework is based on earlier work by Rodriguez et al.~\cite{Rodriguez2009,McCauley2010}, which employed the FDTD method. That approach computed the fields excited by a harmonic expansion of extended sources in an artificial dissipative system, effectively reformulating the dielectric permittivity to deform the complex-frequency integration contour, thereby reducing both the number of simulations and the required simulation time. Instead, our approach uses a physically realistic dissipative system, along with tailored broadband dipole sources positioned at the nodes of a Gaussian quadrature scheme to ensure convergence in the post-processing temporal and surface integration. Compared to standard FDTD methods, which rely on regular staggered grids in both space and time~\cite{Taflove2005}, DGTD achieves higher-order convergence through local polynomial basis functions, adaptive meshing, and efficient time-stepping schemes~\cite{Busch2011}.

This paper is organized as follows. In Sec.~\ref{sec:method}, we outline the theoretical framework for evaluating Casimir interactions within the Maxwell stress tensor approach. Section~\ref{sec:numImp} describes the numerical implementation and its relation to classical scattering problems in the time domain. In Sec.~\ref{sec:applications}, we present two case studies: (i) the Casimir force between two half-spaces, illustrating the temporal integration scheme and its convergence properties; and (ii) the interaction between a finite-size metallic cylindrical object and a half-space, introducing the surface integration scheme necessary for the treatment of systems lacking translational symmetry. Finally, Sec.~\ref{sec:conclusions} summarizes the results and outlines potential future directions.

\section{Maxwell stress tensor approach\label{sec:method}}

We compute fluctuation-induced forces in systems at thermal equilibrium using the Maxwell stress tensor formalism. Specifically, we consider a collection of arbitrarily shaped, nonoverlapping bodies embedded in vacuum at a temperature $T$. The net force on any given object, resulting from quantum and thermal fluctuations of the electromagnetic field, is given by the surface integral of the expectation value of the Maxwell stress tensor operator in the system's thermal state
\begin{equation}
	\mathbf{F} = \oiint_{S} \mathrm{d}S \, \langle\underline{\hat{\mathrm{T}}}(\mathbf{r}, t)\rangle_{T} \cdot \mathbf{n}(\mathbf{r})~.
	\label{eq:cf}
\end{equation}
Here, $S$ denotes any closed surface surrounding the body and $\mathbf{n}(\mathbf{r})$ is the outward-pointing unit vector normal to the surface at position $\mathbf{r}$. In empty space, it is expressed in terms of the quantum correlation functions of the fluctuating electric and magnetic fields, $\hat{\mathbf{E}}(\mathbf{r}, t)$ and $\hat{\mathbf{H}}(\mathbf{r}, t)$, evaluated at the same position $\mathbf{r}$ and time $t$, as follows~\cite{Jackson1975}
\begin{eqnarray}
	\langle \hat{\mathrm{T}}_{ij}(\mathbf{r}, t) \rangle_T &= \epsilon_0 \bigg[\langle \hat{E}_i(\mathbf{r}, t) \hat{E}_j(\mathbf{r}, t) \rangle_T &- \frac{\delta_{ij}}{2} \langle \hat{\mathbf{E}}^2(\mathbf{r}, t) \rangle_T \bigg]\nonumber\\*
	&+ \mu_0 \bigg[\langle \hat{H}_i(\mathbf{r}, t) \hat{H}_j(\mathbf{r}, t) \rangle_T &- \frac{\delta_{ij}}{2} \langle \hat{\mathbf{H}}^2(\mathbf{r}, t) \rangle_T \bigg],\hspace{5mm}
	\label{eq:strTen_quanCorr}
\end{eqnarray}
where $\delta_{ij}$ is the Kronecker delta with $i,j \in \{x, y, z\}$, while $\epsilon_0$ and $\mu_0$ are the vacuum permittivity and permeability, respectively. In thermal equilibrium, the field correlation functions can be evaluated using the fluctuation-dissipation theorem (FDT) as~\cite{Kubo1966,Eckhardt1984}
\begin{subequations}
	\begin{eqnarray}
		\langle \hat{E}_i(\mathbf{r}, t) \hat{E}_j(\mathbf{r}', t) \rangle_T = \frac{\hbar \mu_0}{\pi} \int\limits_{0}^{\infty} &&\mathrm{d}\omega \, \omega^2 \coth\left[\frac{\hbar \omega}{2k_\textrm{B} T}\right] \nonumber\\* &&\times\operatorname{Im} \left\{ G_{ij}^{E}(\mathbf{r}, \mathbf{r}'; \omega) \right\}~,\quad
		\label{eq:quancorr_E} \\
		\langle \hat{H}_i(\mathbf{r}, t) \hat{H}_j(\mathbf{r}', t) \rangle_T = \frac{\hbar \epsilon_0}{\pi} \int\limits_{0}^{\infty} &&\mathrm{d}\omega \, \omega^2 \coth\left[\frac{\hbar \omega}{2k_\textrm{B} T}\right] \nonumber\\* &&\times\operatorname{Im} \left\{ G_{ij}^{H}(\mathbf{r}, \mathbf{r}'; \omega) \right\}~.
		\label{eq:quancorr_B}
	\end{eqnarray}
	\label{eq:quancorr}
\end{subequations}
Although the field operators fluctuate in time, their equal-time correlations are time-independent; therefore, the right-hand side contains no explicit time dependence. Consequently, the expectation value of the stress tensor operator is also time-independent, and the Casimir force in Eq.~\eqref{eq:cf} remains static. These expressions relate the quantum and thermal field fluctuations to the imaginary parts of the electromagnetic Green's tensors $G_{ij}^{\sigma}(\mathbf{r}, \mathbf{r}'; \omega)$ at frequency $\omega$, where $\sigma \in \{E, H\}$ distinguishes between the electric and magnetic contributions. The Green’s tensors encode the geometry and material properties of the system and depend on the observation point $\mathbf{r}$, where the fields are evaluated, and the position $\mathbf{r}'$ of the source that gives rise to scattering in the system. For linear, local, and nonmagnetic objects embedded in vacuum, they satisfy the following inhomogeneous vector Helmholtz equations~\cite{Jackson1975}
\begin{subequations}
	\begin{eqnarray}
		\left[\nabla \times \nabla \times - \frac{\omega^2}{\mathrm{c}^2} \epsilon(\mathbf{r}, \omega)\right]\underline{G}^{E}(\mathbf{r}, \mathbf{r}'; \omega) &=& \delta(\mathbf{r} - \mathbf{r}') \underline{\openone}~,\quad 
		\label{eq:helmeq_gtA}
		\\
		\left[\nabla \times \frac{1}{\epsilon(\mathbf{r}, \omega)} \nabla \times \, - \frac{\omega^2}{\mathrm{c}^2} \right]\underline{G}^{H}(\mathbf{r}, \mathbf{r}'; \omega) &=& \delta(\mathbf{r} - \mathbf{r}') \underline{\openone}~,
		\label{eq:helmeq_gtB}
	\end{eqnarray}
	\label{eq:helmeq_gt}
\end{subequations}
subject to outgoing radiation boundary conditions~\cite{Sommerfeld1949}. Here, $\epsilon(\mathbf{r}, \omega)$ is the piecewise-homogeneous permittivity function, and $\underline{\openone}$ represents the unit dyadic. The total Green’s tensor decomposes as $\underline{G}^{\sigma}(\mathbf{r}, \mathbf{r}'; \omega) = \underline{G}_{0}^{\sigma}(|\mathbf{r} - \mathbf{r}'|; \omega) + \underline{G}^{\sigma}_{\mathcal{S}}(\mathbf{r}, \mathbf{r}'; \omega)$, where $\underline{G}_{0}^{\sigma}$ is the vacuum contribution, which solves Eqs.~\eqref{eq:helmeq_gt} in the isotropic, homogeneous empty vacuum. Although $\underline{G}_0^{\sigma}$ diverges in the coincidence limit $\mathbf{r}' \to \mathbf{r}$, its contribution to Eq.~\eqref{eq:cf} vanishes as it becomes independent of $\mathbf{r}$ and integrates to zero over any closed surface. On the other hand, the scattered Green's tensor $\underline{G}^{\sigma}_\mathcal{S}$ represents the response to a source located at $\mathbf{r}'$ in the presence of material inhomogeneities, such as geometric boundaries and interface reflections. Since $\underline{G}_{0}^{\sigma}$ does not contribute to the net Casimir force, we focus only on $\underline{G}^{\sigma}_\mathcal{S}$ and, for simplicity, omit the subscript $\mathcal{S}$ in the subsequent discussions.

\section{Numerical Implementation\label{sec:numImp}}

The FDT in Eqs.~\eqref{eq:quancorr} connects the expectation value of the Maxwell stress tensor in Eq.~\eqref{eq:strTen_quanCorr} to the imaginary part of the Green's tensors, which capture the electromagnetic response of the system. This relation recasts a quantum electrodynamic effect into a classical electrodynamics problem, governed by Eqs.~\eqref{eq:helmeq_gt}. However, as mentioned above, semi-analytical solutions for these equations exist only for a limited set of highly symmetric geometries. The challenge, therefore, is to develop efficient numerical methods that can evaluate the scattered Green's tensor accurately for arbitrary geometries beyond the reach of semi-analytical methods.

A practical strategy involves interpreting each column of the dyadic Green's tensor as the field response of the system to a point-like dipole source. Specifically, the $ij$-component of the electric Green's tensor, $G_{ij}^{E}(\mathbf{r}, \mathbf{r}'; \omega)$, is proportional to the $i$th component of the electric field at position $\mathbf{r}$ produced by an infinitesimal oscillating electric dipole aligned along the $j$-axis and located at $\mathbf{r}'$~\cite{Tai1994}. The corresponding source current density is expressed by $\mathbf{j}^{E}(\mathbf{r}, \omega) = j^{E}(\omega) \delta(\mathbf{r} - \mathbf{r}') \mathbf{e}_j$, where $j^E(\omega)$ is the frequency-domain amplitude of the dipole source. Analogously, the magnetic Green's tensor, $G_{ij}^{H}(\mathbf{r}, \mathbf{r}'; \omega)$, is obtained by introducing a magnetic dipole. At coincidence, the Green's tensor components follow from the fields as
\begin{equation}
	G^{E}_{ij}(\mathbf{r}, \mathbf{r}; \omega) = \frac{E_i^{(j)}(\mathbf{r}, \omega)}{\mathrm{i}\, \omega \mu_0 j^{E}(\omega)}~, \quad G^{H}_{ij}(\mathbf{r}, \mathbf{r}; \omega) = \frac{H_i^{(j)}(\mathbf{r}, \omega)}{\mathrm{i}\, \omega \epsilon_0 j^{H}(\omega)}~,
	\label{eq:gt}
\end{equation}
where $E_i^{(j)}$ and $H_i^{(j)}$ are the $i$-components of the scattered electric and magnetic fields induced by dipoles oriented along the $j$-axis~\cite{Kristensen2023}. This formulation recasts Casimir-force calculations into a series of classical scattering problems that can be evaluated using standard electromagnetic solvers.

By substituting Eqs.~\eqref{eq:quancorr}, together with Eq.~\eqref{eq:gt}, into Eq.~\eqref{eq:strTen_quanCorr}, the expectation value of the Maxwell stress tensor can be expressed in terms of the scattered electromagnetic fields as
\begin{equation}
	\langle\hat{\mathrm{T}}_{ij}(\mathbf{r}, t)\rangle_T = \frac{\hbar}{\pi} \sum_{\sigma}\operatorname{Im} \left\{ \int\limits_{-\infty}^{\infty} \mathrm{d}\omega\, g_{T}^{\sigma}(\omega)\, \Delta_{ij}^{\sigma}(\mathbf{r}, \omega) \right\}~,
	\label{eq:strTen_freq}
\end{equation}
where the geometry-dependent scattered-field contribution $\Delta_{ij}^\sigma$ is defined as
\begin{equation}
	\Delta_{ij}^{\sigma}(\mathbf{r}, \omega) = \Lambda_{ij}^{\sigma}(\mathbf{r}, \omega) - \frac{\delta_{ij}}{2} \sum_k \Lambda_{kk}^{\sigma}(\mathbf{r}, \omega)~.
	\label{eq:typeFields_freq}
\end{equation}
Here, the Cartesian indices $i,j,k\in\{x,y,z\}$ run over the spatial components and $\sigma\in\{E,H\}$ labels the electric and magnetic contributions
\begin{equation}
	\Lambda_{ij}^{E}(\mathbf{r}, \omega) = \epsilon_0 E_i^{(j)}(\mathbf{r}, \omega)\quad\text{and}\quad
	\Lambda_{ij}^{H}(\mathbf{r}, \omega) = \mu_0 H_i^{(j)}(\mathbf{r}, \omega)~.
\end{equation}
These quantities represent the scattering response of the system. In Eq.~\eqref{eq:strTen_freq}, we introduced the temperature-dependent function
\begin{equation}
	g_{T}^{\sigma}(\omega) = -\mathrm{i} \frac{\omega}{j^\sigma(\omega)} \coth\left[\frac{\hbar \omega}{2k_\textrm{B} T}\right] \Theta(\omega)~,
	\label{eq:kernFunc_freq}
\end{equation}
where $\Theta(\omega)$ is the Heaviside step function. In the limit $\omega \to \infty$, $g_{T}^{\sigma}$ diverges as $\omega^{l+2}$, where $l$ is the order of the first nonvanishing derivative at time $\tau=0$ of the time-domain current response $j^\sigma(\tau)$, whose Fourier transform is $j^\sigma(\omega)$~\cite{Kristensen2023}. Despite its divergent behavior, the imaginary-part operator ``$\operatorname{Im}$'' in Eq.~\eqref{eq:strTen_freq} commutes with the integration, leading to a finite result. This can be justified by interpreting $g_{T}^\sigma$ as a tempered distribution, since it appears only under integration against the smooth, decaying field functions $\Lambda_{ij}^\sigma$.

\subsection{From frequency to time domain}

Direct numerical evaluation of the Maxwell stress tensor through frequency-domain integration along the real axis is hindered by highly oscillatory integrands~\cite{Ford1993}. These oscillations degrade both the accuracy and efficiency of the broadband integral in Eq.~\eqref{eq:strTen_freq}, which accounts for contributions from all fluctuating frequency modes. A standard alternative is contour deformation via Wick rotation, which involves evaluating the Green’s tensor along the imaginary frequency axis. This transformation smooths the integrand, turning it into an exponentially decaying function, thereby ensuring rapid convergence~\cite{Reid2013a}. While frequency-domain methods were among the first numerical approaches for Casimir-type calculations~\cite{Rodriguez2007,Rodriguez2008,Rodriguez2011}, solving Maxwell’s equations at many discrete frequencies renders this method computationally expensive, particularly for complex geometries. As an alternative, time-domain methods have been proposed~\cite{Rodriguez2009,McCauley2010}. These methods excite the system with a short pulse to extract the full spectral response in a single simulation. This efficiency motivates our choice to evaluate Casimir interactions by solving Maxwell’s equations in the time domain.

Following Refs.~\cite{Rodriguez2009,McCauley2010}, the expectation value of the stress tensor can be expressed as a temporal convolution,
\begin{equation}
	\langle \hat{\mathrm{T}}_{ij}(\mathbf{r}, t) \rangle_T = 2\hbar \sum_{\sigma} \int\limits_{0}^{\infty} \mathrm{d}\tau\, \operatorname{Im} \left\{ g_{T}^{\sigma}(-\tau) \right\} \Delta_{ij}^{\sigma}(\mathbf{r}, \tau)~.
	\label{eq:strTen_time}
\end{equation}
Within the proposed framework, the evaluation of the averaged Maxwell stress tensor reduces to calculating the temporal kernel $\operatorname{Im}\{g_{T}^{\sigma}(-\tau)\}$ and computing the scattered electromagnetic response $\Delta_{ij}^\sigma(\mathbf{r},\tau)$. The former admits an analytic treatment, while the latter is obtained by numerically solving the time-dependent Maxwell curl equations using our own implementation of the DGTD method. It is worth noting that the numerical scheme is general and can be applied to other numerical electromagnetic field solvers in the time domain~\cite{Rodriguez2009,McCauley2010,Kilianski2025}. Moreover, the scattered response $\Delta_{ij}^{\sigma}$ is temperature-independent, allowing thermal effects to be treated through the kernel without additional computational overhead.

\subsubsection{Source current density and temporal kernel}

The behavior of both the kernel and the scattered response is tied to the temporal profile of the source current density. This is formally rather arbitrary within the above-described approach, provided it is causal, continuous, and produces divergence-free fields. Causality and physical realizability require the current density to be real and vanish for $\tau < 0$, with $\tau = 0$ marking the excitation onset. In addition, to ensure consistency with Maxwell’s curl equations and a divergence-free evolution, the current density and its first two derivatives must vanish at $\tau = 0$~\cite{Kristensen2023,Taflove2005}.

In practice, a specific form for the current profile is chosen to accelerate the convergence of the temporal integral in Eq.~\eqref{eq:strTen_time}. Building on and extending the approach in Ref.~\cite{Kristensen2023}, we have found an exponentially damped polynomial current of the form,
\begin{equation}
	j^\sigma(\tau) = j_0^\sigma (\gamma \tau)^{s-1} \left[s - \chi^\sigma \gamma \tau\right] \mathrm{e}^{-\gamma \tau} \Theta(\tau)~,
	\label{eq:currDens_time}
\end{equation}
where the integer $s \geq 4$ ensures that both the current and its first two time derivatives vanish at $\tau = 0$. The amplitude is set by $j_0^\sigma$, and causality is enforced by the Heaviside step function $\Theta(\tau)$. The parameter $\gamma > 0$ controls the decay rate and thereby the spectral bandwidth of the excitation, while the discrete parameter
\begin{equation}
	\chi^\sigma =
	\begin{cases}
		1 & \text{if } \sigma = E~, \\
		0 & \text{if } \sigma = H~,
	\end{cases}
\end{equation}
distinguishes electric ($\sigma=E$) from magnetic ($\sigma=H$) dipole excitations, modifying the pulse shape. In practice, the choice of current density in Eq.~\eqref{eq:currDens_time}, with $s = 4$ and $\gamma \approx \mathrm{c} / 2L$, where $L$ is the minimum surface-to-surface separation between bodies (cf. Fig.~\ref{fig:2hs_hs-cyl_mesh}), ensures proper temporal convergence for both electric and magnetic dipole excitations at zero and finite temperature, cf. Appendix~\ref{app:sourceParam}. Indeed, in contrast to Ref.~\cite{Kristensen2023}, where only the electric field is relevant, the slow decay of the scattered magnetic field, due to the existence of long-lived nonoscillatory modes~\cite{Intravaia2009,Intravaia2010,Henkel2010}, affects the system's electromagnetic response at long times.

The kernel $\operatorname{Im}\left\{g_{T}^{\sigma}(-\tau)\right\}$, as defined in Eq.~\eqref{eq:kernFunc_freq}, depends on the Fourier-transformed current density,
\begin{equation}
	j^\sigma(\omega) = j_0^\sigma s!\gamma^{s-1}\frac{[(1 - \chi^\sigma)\gamma - \mathrm{i}\omega]}{(\gamma - \mathrm{i}\omega)^{s+1}}~.
	\label{eq:currDens_freq}
\end{equation}
Understanding this connection is key to assessing the convergence of the time-domain integral in Eq.~\eqref{eq:strTen_time}. A rapid decay of the kernel, combined with the intrinsic damping of electromagnetic fields in dissipative media, facilitates numerical convergence. This, in turn, shortens the simulation time required for a given accuracy. The full time dependence of $\operatorname{Im}\{g_{T}^{\sigma}(-\tau)\}$ is obtained by Fourier transforming Eq.~\eqref{eq:kernFunc_freq} through contour integration in the complex-frequency plane, with $\omega \rightarrow \zeta = \omega + \mathrm{i}\xi$~\cite{Kristensen2023}. Due to causality, $j^\sigma(\zeta)$ is analytic in the upper half-plane and satisfies the Schwarz reflection principle $[j^\sigma(\zeta)]^* = j^\sigma(-\zeta^*)$, as required for the Fourier transform of any real-valued time-domain function. Therefore, the only term contributing with single poles to Eq.~\eqref{eq:kernFunc_freq} when closing the contour in the upper half-plane is the thermal factor $\coth[\hbar\omega/2k_\textrm{B} T]$. These are located at the Matsubara frequencies $\zeta_m = \mathrm{i}\xi_m = \mathrm{i} m \omega_{\rm th}$, with $\omega_{\rm th} = 2\pi k_\textrm{B} T/\hbar$. Applying the residue theorem yields
\begin{equation}
	\operatorname{Im}\left\{g_{T}^{\sigma}(-\tau)\right\} = \frac{\omega_{\rm th}}{2\pi} \sideset{}{'}\sum_{m=0}^\infty \frac{\xi_m}{j^\sigma(\mathrm{i}\xi_m)} \mathrm{e}^{-\xi_m \tau}~,
	\label{eq:img_currDens}
\end{equation}
where the prime indicates that the $m = 0$ term is weighted by a factor $1/2$. For the current density in Eq.~\eqref{eq:currDens_freq} and $\sigma \in \{E, H\}$, the Matsubara sum in Eq.~\eqref{eq:img_currDens} admits a closed-form representation as
\begin{widetext}
	\begin{equation}
		\operatorname{Im}\left\{g_{T}^{\sigma}(-\tau)\right\} = \tilde{g}_{0}^\sigma 
		\begin{cases}
			\displaystyle \frac{\omega_{\rm th}}{2\gamma}+\sum_{l=0}^{s+1} \binom{s+1}{l} \left( \frac{\omega_{\rm th}}{\gamma} \right)^{l+1} \operatorname{Li}_{-l}(\mathrm{e}^{-\omega_{\rm th} \tau}) & \sigma = E~, \\[12pt]
			\displaystyle\sum_{l=0}^{s} \binom{s}{l} \left( \frac{\omega_{\rm th}}{\gamma} \right)^{l+2} \operatorname{Li}_{-(l+1)}(\mathrm{e}^{-\omega_{\rm th} \tau}) & \sigma = H~,
		\end{cases}
		\label{eq:img_time}
	\end{equation}
\end{widetext}
where $\tilde{g}_{0}^\sigma = \gamma^3 / 2\pi j_0^\sigma s!$ and $\operatorname{Li}_{l}(z)$ is the polylogarithm function of order $l$~\cite{Wood1992}.

\subsubsection{The DGTD method\label{subsubsec:dgtd}}

To solve Maxwell’s equations numerically for arbitrary scattering configurations, we employ the nodal DGTD method in the form introduced by Hesthaven and Warburton~\cite{Hesthaven2002}. This finite-element-based approach is designed for time-domain integration of partial differential equations in flux-conservative form~\cite{Hesthaven2008}. In this method, the computational domain is discretized using an unstructured mesh composed of finite elements that fully cover the simulation region. Within each element, the electromagnetic fields are expanded in a set of Lagrange polynomials of order $p$, which act as local basis functions. The expansion coefficients are related to the field amplitudes at the corresponding nodal points~\cite{Busch2011}. This local expansion promotes memory efficiency and parallelization~\cite{Busch2011}, thereby reducing computational cost for large-scale three-dimensional simulations. A global solution is then constructed by introducing the so-called upwind numerical flux at element interfaces, ensuring inter-element coupling and enforcement of boundary conditions~\cite{Hesthaven2002}. Through spatial discretization, Maxwell’s curl equations are cast into a semi-discrete form, i.e., a set of first-order coupled ordinary differential equations for the time-dependent expansion coefficients. These are integrated in time using an explicit fourth-order low-storage Runge--Kutta scheme with 14 stages~\cite{Niegemann2012}, yielding a flexible, high-order time-stepping technique.

In the time domain, frequency dispersion is incorporated through auxiliary differential equations (ADEs), which introduce auxiliary currents to model material polarization. These ADEs are solved alongside Maxwell’s equations in the time-stepping scheme. Furthermore, to inject dipolar excitations and evaluate the scattered field response, we use the Total-Field/Scattered-Field (TF/SF) technique. This approach exploits the linearity of Maxwell’s equations to decompose the fields into incident and scattered components by splitting the computational domain into total-field (TF) and scattered-field (SF) regions. The DGTD framework is particularly well suited to this formulation, as field values are stored independently on either side of the element interfaces. Consequently, the analytically known source-induced incident field distribution can be directly added or subtracted at the TF/SF boundary. Notably, singularities from localized point sources are inherently regularized by placing the source within the SF region, thereby aligning with the scattered Green’s tensor formalism introduced earlier.

The spatial convergence of the scheme depends on both the average edge length, $h$, and the polynomial order, $p$. Increasing $p$ enhances spatial resolution by adding more interpolation nodes within each element. In this work, we use straight-sided tetrahedra, which are particularly convenient for resolving multiscale features and handling curved geometries more naturally than regular grids, reducing staircasing artifacts. When the geometry is well-represented by the mesh, i.e., it is discretization-conforming, the method achieves high-order convergence of $\mathcal{O}(h^{p+1})$~\cite{Busch2011}. However, for non-conforming meshes with unresolved curvature, convergence may be limited to $\mathcal{O}(h^2)$~\cite{Viquerat2015}. For smoothly varying boundaries, curvilinear elements can, in principle, be used to restore high-order convergence~\cite{Busch2011,Viquerat2015}.

\section{Applications\label{sec:applications}}

In this section, we present the numerical results of the proposed scheme, focusing on the Casimir force between two objects. A key step in the procedure is the closed-surface integration in Eq.~\eqref{eq:cf}. In principle, this requires evaluating the stress tensor at multiple points of the closed surface $S$, leading to a large number of time-domain simulations—six per evaluation point, corresponding to three orthogonal dipole orientations and two source types. Although these simulations can be easily parallelized, the computational cost can become prohibitive at high spatial resolutions, even on computing clusters. The closed surface is chosen to enclose the body on which we want to calculate the force. While the geometry of $S$ is theoretically arbitrary, certain choices can be numerically more efficient than others. For example, if one body is enclosed by a cube, the technical evaluation of the integral simplifies considerably. In the limit of an infinitely large cube, the only contributing surface is the face between the two bodies, $S_{0}$, since the amplitudes of the scattered field on the remaining faces decay rapidly with distance and their contributions vanish at infinity. The essential physics is then fully captured by integrating only over the infinitely extended surface $S_{0}$. Further simplifications can rely on the symmetries of specific systems.

We now illustrate this procedure through two representative examples. As a benchmark, we first consider the Casimir interaction between two metallic half-spaces separated by a vacuum gap of width $L$. This setup provides insight into the temporal integration in Eq.~\eqref{eq:strTen_time}. In addition, this geometry offers computational simplicity and admits a semi-analytical treatment using the Lifshitz formula~\cite{Lifshitz1956}, which provides an independent reference calculation for assessing the accuracy and convergence of the numerical approach. As a second example, we consider a finite-size metallic cylindrical object positioned above a half-space, with its axis perpendicular to the surface. This less-symmetric configuration underscores the surface-integration scheme that links the stress tensor to the force. Furthermore, it illustrates the adaptability of the approach in accommodating arbitrary three-dimensional geometries, including scenarios where the frequently employed PFA proves ineffective.

\subsection{Casimir force between half-spaces\label{subsec:transSym}}

We consider two identical, metallic semi-infinite planes separated by a distance $L$ along the $z$-axis, cf. Fig.~\ref{fig:2hs_hs-cyl_mesh}(a). The optical response of the metal is modeled as that of silver, described by a standard Drude dielectric function
\begin{equation}
	\epsilon(\omega)=1-\frac{\omega_p^2}{\omega(\omega + \mathrm{i}\Gamma)}~,
	\label{eq:drude}
\end{equation}
with plasma frequency $\omega_p = 1.39\times10^{16}\, \textrm{s}^{-1}$ ($\hbar\omega_p = 9\, \textrm{eV}$) and damping rate $\Gamma = 3.23\times10^{13}\, \textrm{s}^{-1}$ ($\hbar\Gamma = 21\, \textrm{meV}$)~\cite{Johnson1972}. Due to the translational invariance in the $xy$-plane, we choose the integration surface $S_{0}$ to lie at the mid-plane at $z = L/2$, equidistant from both interfaces. This surface extends infinitely in the $x$- and $y$-directions and has a unit normal vector $\mathbf{n} = -\mathbf{e}_z$. The symmetry of the system ensures that the net Casimir force is directed along the $z$-axis, i.e., $\mathbf{F}=F_{z}\mathbf{e}_z$. Accordingly, the relevant component of the Maxwell stress tensor is $\langle \hat{\mathrm{T}}_{zz}(\mathbf{r}, t) \rangle_T$, which needs to be integrated over the in-plane coordinates $\mathbf{r}_\parallel = (x,y)$ lying in the plane $z=L/2$. However, due to lateral invariance, $\langle \hat{\mathrm{T}}_{zz}(\mathbf{r}, t) \rangle_T$ is spatially uniform over $z=L/2$. The Casimir force per unit area can therefore be computed at a single spatial position, say $\mathbf{r}_{0}=(0,0,L/2)$, as
\begin{equation}
\label{eq:tensor}
	\frac{F_{z}}{S_{0}} = - \langle \hat{\mathrm{T}}_{zz}(\mathbf{r}_{0}, t) \rangle_T~,
\end{equation}
thus reducing the numerical problem to a single time integration.

As mentioned above, the symmetry of the system enables a semi-analytical solution, and Eq.~\eqref{eq:tensor} can be recast in terms of the Lifshitz formula~\cite{Lifshitz1956,Dalvit2011}
\begin{equation}
	\frac{F_{z}}{S_{0}}=-k_\textrm{B}T
	\sideset{}{'}\sum_{m=0}^{\infty}
	\int\frac{\mathrm{d}^{2}\mathbf{q}}{(2\pi)^{2}}\sum_{u
	}
	2\kappa\frac{\left[r_{u}(\mathrm{i}\xi_{m},q)\mathrm{e}^{-\kappa L}\right]^{2}}{1-\left[r_{u}(\mathrm{i}\xi_{m},q)\mathrm{e}^{-\kappa L}\right]^{2}}~,
	\label{eq:LifshitzFormula}
\end{equation}
which describes the pressure on one of the half-spaces in terms of the component of the electromagnetic wave-vector parallel to the planes, $\mathbf{q}$, with modulus $q=|\mathbf{q}|$, and the orthogonal wave-vector component in vacuum, $\kappa=\sqrt{q^{2}+\xi_{m}^{2}/\mathrm{c}^{2}}$, evaluated at the imaginary Matsubara frequencies. The reflection coefficients $r_{u}$, with $u\in\{\mathrm{TE},\mathrm{TM}\}$, describe the scattering of the fields at the half-space. Within the present context, $r_{u}$ is given using the Fresnel formulae~\cite{Jackson1975,Intravaia2022}. As above, the prime in the sum indicates that the $m=0$ term carries a weight $1/2$. For the configuration considered here, Eq.~\eqref{eq:LifshitzFormula} predicts that the two half-spaces attract each other~\cite{Lifshitz1956}. 

In terms of the Lifshitz formula, the interaction can be characterized by the relative size of $L$ with respect to two characteristic length scales: the plasma wavelength $\lambda_{p}=2\pi \mathrm{c}/\omega_{p}$, which is approximately $136\,\textrm{nm}$ for silver, and the thermal or Wien wavelength $\lambda_{\rm th} = 2\pi \mathrm{c}/\omega_{\rm th}=\hbar \mathrm{c} / k_\textrm{B} T$, which roughly corresponds to the characteristic length scale in Wien's displacement law and is on the order of $7.6\, \mu\textrm{m}$ at room temperature. For $L \ll \lambda_p$, the interaction is in the nonretarded regime, where the effects of retardation can be neglected. In this limit, Eq.~\eqref{eq:LifshitzFormula} becomes~\cite{Henkel2004,Intravaia2022}
\begin{equation}
	\begin{aligned}
		\frac{F_{z}}{S_{0}}&\sim-\frac{\hbar}{8\pi^{2}L^{3}}\int\limits_{0}^{\infty}\mathrm{d}\xi\, \operatorname{Li}_{3}\left(\left[\frac{\epsilon(\mathrm{i}\xi)-1}{\epsilon(\mathrm{i}\xi)+1}\right]^{2}\right)\\
		&\stackrel{\Gamma/\omega_{p}\ll 1}{\sim}
		-\frac{\hbar \mathrm{c}\pi^{2}}{240L^{4}}\left[ \eta\frac{L}{\lambda_{p}}-\frac{15\zeta(3)}{\pi^4}\frac{\Gamma L}{\mathrm{c}}\right]~,
	\end{aligned}
	\label{eq:asympLifshitzFormula}
\end{equation}
where $\eta\approx1.193$ and $\zeta(x)$ is the Riemann zeta function, for which $\zeta(3)\approx1.202$. Note that the previous expression does not depend on the temperature of the system, indicating that, in this regime, the quantum fluctuations dominate over the thermal fluctuations. For separations $L\gg \lambda_{p}$, retardation effects become significant and, if the temperature is kept at zero, the Lifshitz formula recovers Casimir's original result~\cite{Casimir1948a}, given by
\begin{equation}
\label{eq:CasimirP}
\frac{F_{z}}{S_{0}}\sim -\frac{\hbar \mathrm{c} \pi^{2}}{240 L^{4}}
\end{equation}
for the attraction between two perfectly reflecting plates. In this limit, the force loses all dependence on the material parameters, which remain relevant only through the condition $L\gg \lambda_{p}$. Corrections due to finite temperature become appreciable when $L \gtrsim \lambda_{\rm th}$. In the limit $L \gg \lambda_{\rm th}$, the classical thermal fluctuations dominate the interaction and the Lifshitz formula tends to
\begin{equation}
\label{eq:asympthermal}
\frac{F_{z}}{S_{0}}\sim -\frac{\zeta(3)}{8\pi}\frac{k_\textrm{B}T}{L^{3}}~.
\end{equation}
The above expressions and associated quantities serve as a benchmark for our numerical computations.

\subsubsection{Time integration\label{subsubsec:timeint}}

Due to the symmetry of the system, we have carried out the computations within a cylindrical computational domain centered at $\mathbf{r}_{0}$, with radius $R$ and height $2H + L$, terminated with Silver--M\"uller boundary conditions~\cite{Busch2011}. The metallic half-spaces are approximated as finite disks of radius $R$ and thickness $H$, forming the TF region, cf. Fig.~\ref{fig:2hs_hs-cyl_mesh}(a). Due to the dissipative nature of metals, the thickness $H$ is chosen such that contributions from the opposite interfaces can be neglected, effectively recovering the configuration of two facing half-spaces originally considered by Lifshitz~\cite{Lifshitz1956}. The radius $R$ is chosen to minimize spurious back-reflections from the edge of the computational domain reaching $\mathbf{r}_{0}$~\cite{Kristensen2023}. Within the metallic disks, Maxwell’s equations are solved in conjunction with the ADE for the Drude model~\cite{Busch2011}. The SF region between the disks is filled with air, and the TF/SF boundaries coincide with the metal--vacuum interfaces. The tetrahedral mesh, generated by Gmsh~\cite{Geuzaine2009}, is locally refined near the evaluation position $\mathbf{r}_{0}$ with an average edge length of $h = L / 10$ within a cylindrical region of radius $L$. The mesh is linearly coarsened towards the boundaries of the computational domain, reaching $h = L$ at the outermost limits. We use a global polynomial order $p = 3$, ensuring a resolution of up to two interpolation nodes per smallest relevant wavelength~\cite{Taflove2005,Busch2011}. For an effective numerical treatment of Maxwell’s equations, it is also advantageous to use dimensionless units. This approach entails scaling all fields and lengths with a reference field strength $E_{0}$ and a length scale $\ell_{0}$. For the application considered here, we use $E_{0}=1\,\textrm{V/m}$ and a length scale $\ell_{0}=1\,\textrm{nm}$.

\begin{figure*}[ht]
	\vspace{0.05cm}
	\includegraphics[width=0.98\linewidth]{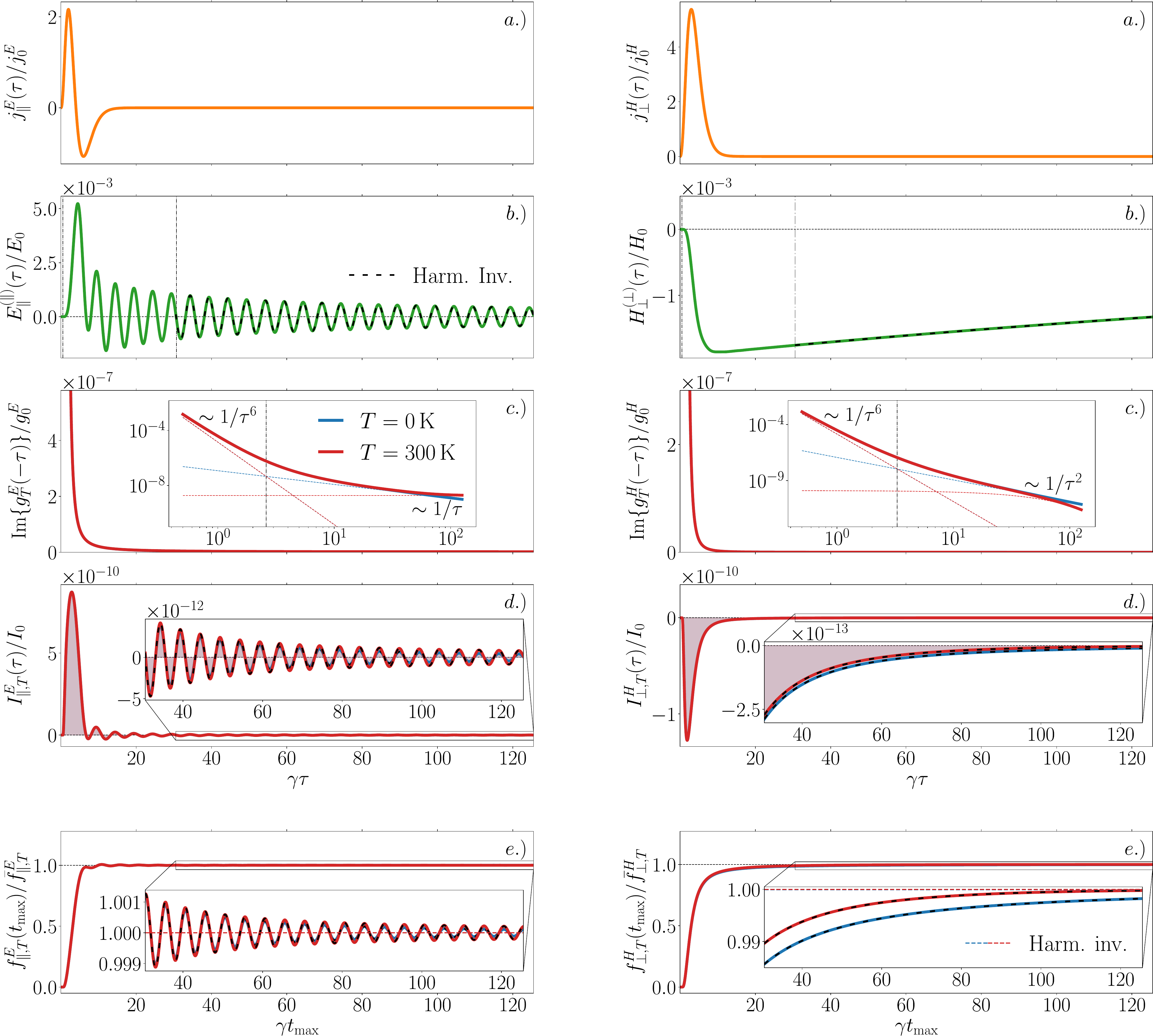}
	\begin{tikzpicture}[remember picture, overlay]
        \node at (-13.2,16.0) {{\bf Electric dipole source}};
    	\node at (-3.85,16.0) {{\bf Magnetic dipole source}};
  	\end{tikzpicture}
	\caption{{\bf Left column:} Relevant quantities characterizing the time-integration scheme for two silver half-spaces separated by $L = 20\,\mathrm{nm}$, excited by an electric dipole oriented parallel to the planes. All quantities are scaled with a reference field strength $E_{0}=1\,\textrm{V/m}$ and a length scale $\ell_{0}=1\,\textrm{nm}$. Each panel depicts a different component of the scheme: (a) Temporal profile of the current density, as defined in Eq.~\eqref{eq:currDens_time} for $s=4$, expressed in units of $j_0^E = \epsilon_0 E_0 \ell_{0}^2 \mathrm{c}$; (b) Scattered electric field $E_{\parallel}^{(\parallel)}$ at $\mathbf{r}_{0}=(0,0,L/2)$ in units of $E_0$; (c) Kernel function in units of $g_0^E = \mathrm{c}^3 / \ell_{0}^3 j_0^E$, with a logarithmic scale inset showing the behavior at zero (blue) and room temperature (red). For details on the asymptotic behavior refer to Appendix~\ref{app:asym}; (d) Time integrand $I_{\parallel, T}^{E}(\tau)=\operatorname{Im}\{g_T^E(-\tau)\} \epsilon_{0}E_{\parallel}^{(\parallel)}(\mathbf{r}_{0}, \tau)$ in units of $I_0 = g_0^E \epsilon_{0}E_0 = \mathrm{c}^2 / \ell_0^5$, with an inset focusing on the long-time behavior; (e) Numerically computed cumulative integral as a function of the simulation time $t_{\mathrm{max}}$, normalized to the reference value $\bar{f}_{\parallel, T}^{E}$, obtained via complex frequency-domain integration of the corresponding Green's tensor component, cf. Appendix~\ref{app:RefValue}. The dashed lines correspond to the result incorporating long-time contributions. {\bf Right column:} Same as the left column, but for a perpendicularly oriented magnetic dipole. All quantities are expressed in the following units: (a) $j_0^H = E_0 \ell_{0}^2$; (b) $H_0 = E_0\sqrt{\epsilon_0 / \mu_0}$; (c) $g_0^H = \mathrm{c}^3 / \ell_{0}^3 j_0^H$; (d) $I_0 = \mathrm{c}^2 / \ell_0^5$.}
	\label{fig:timeint_2hs_f}
\end{figure*}
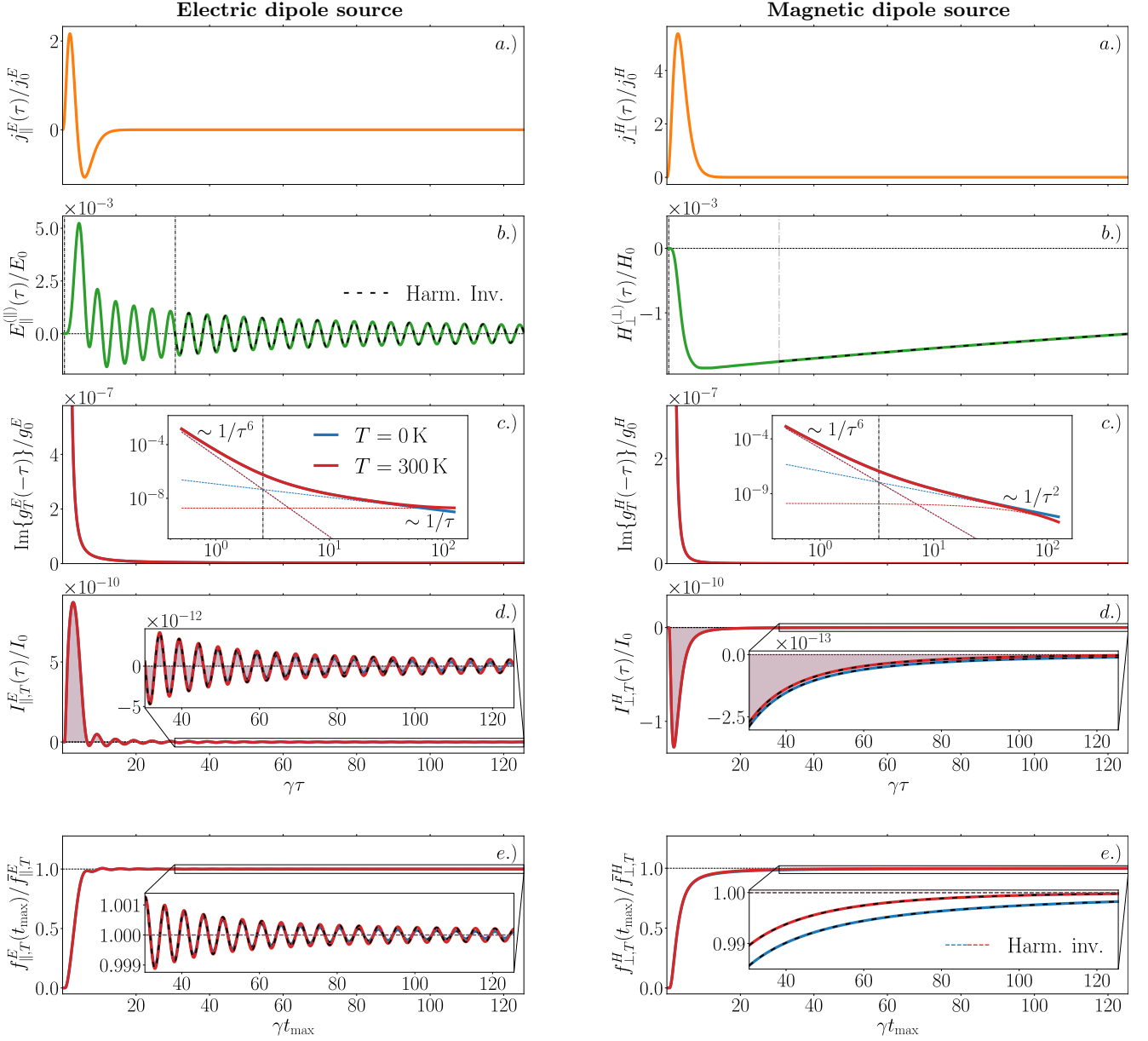

The excitation pulse is located at $\mathbf{r}= \mathbf{r}_{0}$ and introduced into the simulation scheme through the TF/SF technique. Azimuthal invariance implies that at $\mathbf{r}_{0}$, we need only distinguish between $j=\parallel$ for the direction parallel to the planes and $j=\perp$ for the direction perpendicular to them. In panels (a) of Fig.~\ref{fig:timeint_2hs_f}, we illustrate the temporal behavior for a parallel-oriented electric dipole source (left column) and a perpendicular-oriented magnetic dipole source (right column). The pulse profile corresponds to the definition in Eq.~\eqref{eq:currDens_time}, where for both the electric and magnetic dipoles we used $s=4$~\cite{Kristensen2023}. The pulse decays at a rate $\gamma$, a tunable parameter typically set to $\gamma \approx \mathrm{c} / 2L$, which controls the spectral bandwidth, as discussed in Appendix~\ref{app:sourceParam}.

In panels (b) of Fig.~\ref{fig:timeint_2hs_f}, we illustrate in the left and right columns the time evolution of the corresponding scattered fields recorded directly at the nodal point $\mathbf{r} = \mathbf{r}_{0}$, without spatial interpolation. Due to the symmetry of the configuration, the scattered fields $E_i^{(j)}(\mathbf{r}_{0},\tau)$ and $H_i^{(j)}(\mathbf{r}_{0},\tau)$ vanish for $i\neq j$. This property is inherited by the related quantities defined below. Note that in panels (b) of Fig.~\ref{fig:timeint_2hs_f}, the field amplitudes remain zero for $\tau \leq L / \mathrm{c}$, reflecting the roundtrip time for the excitation to reach the surface of the metal and return. For $\tau \geq L / \mathrm{c}$, transient reflections appear, followed by a decay regime once the initial pulse has sufficiently faded. In the source-free regime for the specific configuration considered, the electric-field amplitude in the left column of Figs.~\ref{fig:timeint_2hs_f}(b) exhibits exponentially damped oscillations at a frequency close to the surface-plasmon resonance $\omega_{sp}\approx\omega_p/\sqrt{2}$, corresponding to the dominant pole of the nonretarded electric Green’s tensor~\cite{Dalvit2011}. On the other hand, the amplitude of the magnetic field in the right column of Figs.~\ref{fig:timeint_2hs_f}(b) decays exponentially without oscillatory behavior.

To assess the temporal convergence of Eq.~\eqref{eq:strTen_time}, we define the component-wise contributions to the local Casimir force per unit area as
\begin{equation}
	f_{ij, T}^{\sigma}(\mathbf{r}) = 2\hbar \int\limits_{0}^{t_{\textrm{max}}\to\infty} \mathrm{d}\tau\, \operatorname{Im} \left\{ g_{T}^{\sigma}(-\tau) \right\} \Lambda_{ij}^{\sigma}(\mathbf{r}, \tau)~.
	\label{eq:f_time}
\end{equation}
Numerical evaluation of Eq.~\eqref{eq:f_time} requires truncating the upper limit at $t_{\textrm{max}}$, introducing a trade-off between computational efficiency and numerical accuracy. In Ref.~\cite{Rodriguez2009}, $t_{\textrm{max}}$ is chosen such that the field amplitude decays below a specified error threshold in an artificially dissipative medium with tunable dissipation. However, extending this criterion to physically realistic systems can be computationally demanding and may result in prohibitively long simulation times, particularly for cavities and resonators~\cite{Niegemann2009}. To reduce computation time while maintaining accuracy, we instead choose $t_\textrm{max}$ to capture only the relevant field information based on the system's physical characteristics~\cite{Niegemann2009,Govindjee2012}. The long-time response is then extrapolated by fitting the finite-duration signal over the interval $[t_0, t_\mathrm{max}]$ to a linear superposition of exponentially damped oscillatory modes, where $t_0$ is chosen such that the initial pulse has sufficiently decayed, e.g., $\gamma t_0 = 30$ in Figs.~\ref{fig:timeint_2hs_f}(b). For this purpose, we use a time-domain harmonic inversion technique based on the filter-diagonalization method~\cite{Mandelshtam1997,Wall1995}, which has the advantage of providing an analytical expression for the long-time response field. The modal parameters are extracted using the Harminv implementation within a finite frequency window~\cite{Johnson2004}. In the left and right columns of Figs.~\ref{fig:timeint_2hs_f}(b), we illustrate the excellent agreement between reconstructed and simulated fields for $\gamma\tau\geq 30$. Therefore, to account for contributions beyond the simulation window, long-time corrections are obtained analytically by effectively integrating Eq.~\eqref{eq:f_time} from $t_\textrm{max}$ to infinity using the harmonic inversion parameters extracted from the simulated data, i.e.,
\begin{equation}
	\begin{aligned}
		f_{ij, T}^{\sigma}(\mathbf{r}) &= 2\hbar \left[\int\limits_{0}^{t_{\textrm{max}}} \mathrm{d}\tau\, \operatorname{Im} \left\{ g_{T}^{\sigma}(-\tau) \right\} 
		\Lambda_{ij}^{\sigma}(\mathbf{r}, \tau)\right.
		\\
		&\left.+ \sum_n\operatorname{Re}\left\{d_n \int\limits_{t_{\textrm{max}}}^{\infty} \mathrm{d}\tau\, \operatorname{Im} \left\{ g_{T}^{\sigma}(-\tau) \right\} 
		\mathrm{e}^{-\mathrm{i}a_n\tau}\right\}\right]~.
	\end{aligned}
	\label{eq:f_time-HI}
\end{equation}
The first integral on the right-hand side of Eq.~\eqref{eq:f_time-HI} is evaluated using standard quadrature techniques, with $\Lambda_{ij}^{\sigma}$ the simulated scattered fields, while the second is performed analytically with $d_n=A_n\mathrm{e}^{\mathrm{i}\phi_n}$ and $a_n=\omega_n-\mathrm{i}\alpha_n$ the harmonic inversion parameters~\cite{Johnson2004}. Here, $A_n$ is the mode amplitude, $\phi_n$ the phase shift, $\omega_n$ the resonant frequency, and $\alpha_n>0$ the decay constant of the $n$th mode. The combination of direct numerical integration and the semi-analytical long-time correction improves convergence compared to an approach based only on the first term of Eq.~\eqref{eq:f_time-HI}. Although the harmonic inversion semi-analytically captures the long-time dynamics, it remains a numerical technique that introduces approximation errors. Consequently, the accuracy of Eq.~\eqref{eq:f_time-HI} still depends on $t_{\textrm{max}}$; therefore, a sufficiently rapid decay of the integrands remains essential to minimize errors. Besides field dissipation, convergence is also governed by the behavior of the kernel $\operatorname{Im}\{g_{T}^{\sigma}(-\tau)\}$, defined in Eq.~\eqref{eq:img_time} and illustrated in the left and right columns of Figs.~\ref{fig:timeint_2hs_f}(c) for $T = 0\,\textrm{K}$ and $T = 300\,\textrm{K}$. As shown in Figs.~\ref{fig:timeint_2hs_f}(d) for both the electric and magnetic cases, the kernel amplifies short-time field contributions while suppressing those at longer times. Notably, the divergence at $\tau = 0$ is physically regularized by causality, since the scattered field vanishes for $\tau < L / \mathrm{c}$. 

\begin{figure}
	\includegraphics[width=\linewidth]{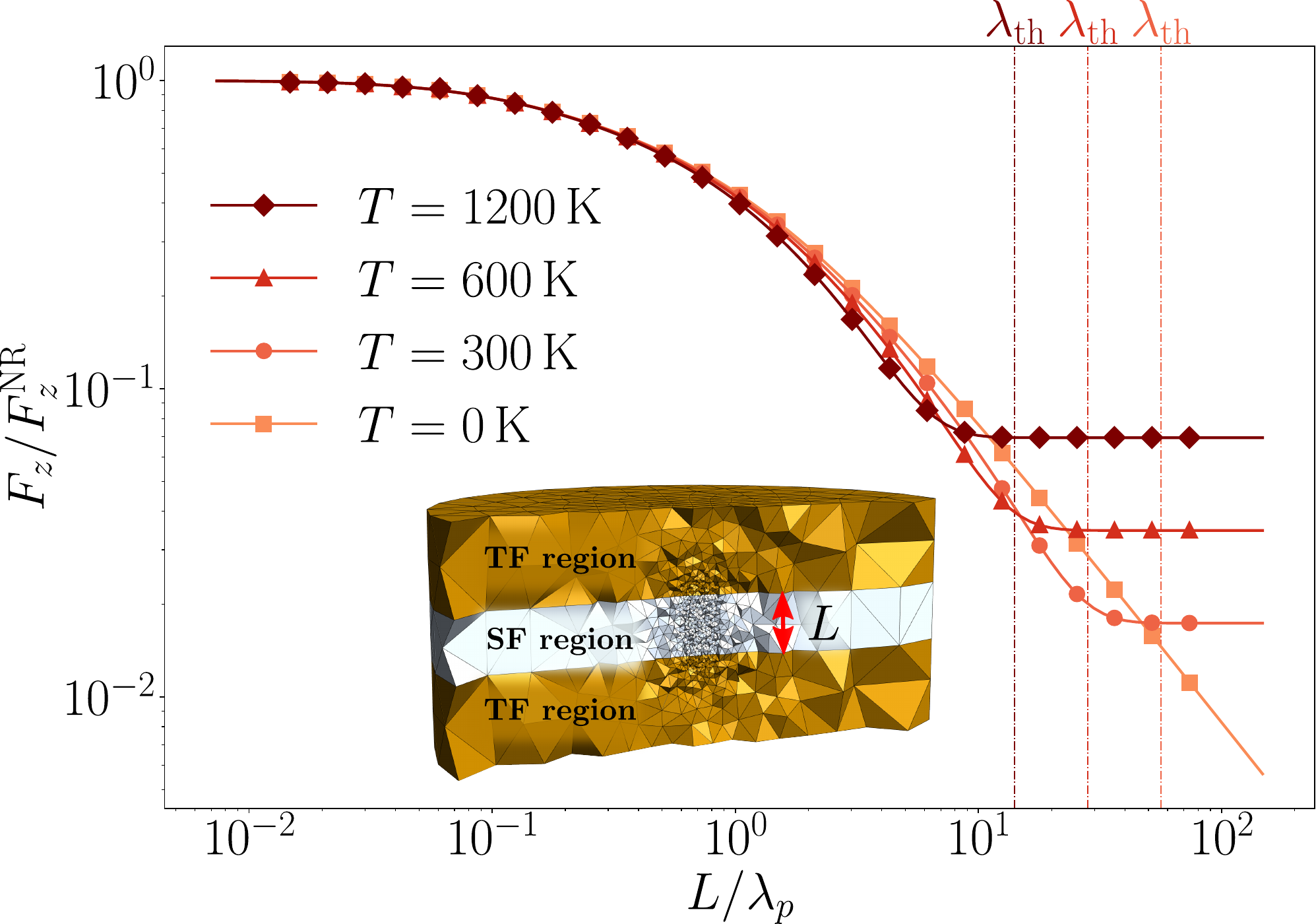}
	\caption{Casimir force between two half-spaces as a function of their separation $L$. The force is normalized to the nonretarded asymptotic value, cf. Eq.~\eqref{eq:asympLifshitzFormula}. Symbols denote numerical results, while solid lines represent semi-analytical reference calculations based on the Lifshitz formula in Eq.~\eqref{eq:LifshitzFormula}. At short separations, the numerical results recover the $\sim -1 / L^{3}$ behavior predicted by Eq.~\eqref{eq:asympLifshitzFormula} and are independent of temperature, whereas at large separations they reproduce the $\sim -1 / L^{4}$ and $\sim -1 / L^{3}$ dependencies expected from Eqs.~\eqref{eq:CasimirP} and \eqref{eq:asympthermal} at zero and finite temperature, respectively. The inset shows a representative mesh of the plane--plane geometry generated with Gmsh~\cite{Geuzaine2009}.
	\label{fig:cf_2hs_L_T}}
\end{figure}

Inserting the numerical results from Eq.~\eqref{eq:f_time-HI} into the $zz$ component of the Maxwell stress tensor in Eq.~\eqref{eq:strTen_time} yields the Casimir force per unit area along the $z$ direction. Figure~\ref{fig:cf_2hs_L_T} shows the numerical results for the force as a function of the separation $L$, ranging from $1\,\textrm{nm}$ to $20\,\mu\textrm{m}$, normalized to the asymptotic value of the Lifshitz formula in Eq.~\eqref{eq:asympLifshitzFormula}. The outcome of our numerical procedure (symbols) agrees very well with the semi-analytical results in Eq.~\eqref{eq:LifshitzFormula} (solid lines) across a wide range of separations and temperatures. When retardation is negligible ($L\ll\lambda_{p}$), the numerical data approach unity, become insensitive to temperature, and reproduce the expected asymptotic scaling $F_z \sim -1/L^3$ predicted by Eq.~\eqref{eq:asympLifshitzFormula}. At larger separations, two different trends are observed. First, when the finite speed of light becomes relevant~\cite{Intravaia2022,Buhmann2012}, the force tends toward Casimir's original expression in Eq.~\eqref{eq:CasimirP}, approaching the scaling $\sim -1/L^4$. In this region, the contribution of the thermal fluctuations is still sufficiently small, although deviations are visible. When $L\gg \lambda_{\rm th}$, thermal effects become significant and, in accordance with Eq.~\eqref{eq:asympthermal}, the data progressively recover the expected scaling of $\sim -1/L^{3}$. The temperature dependence of the prefactor in Eq.~\eqref{eq:asympthermal} is distinctly visible within the normalization scheme used in Fig.~\ref{fig:cf_2hs_L_T}.

\subsubsection{Convergence studies\label{subsubsec:convstud}}

To quantify the accuracy and convergence properties of the numerical scheme, we evaluate the relative error of the Casimir force per unit area, computed independently for each component as
\begin{equation}
	\left|\frac{\Delta f_{ij,T}^{\sigma}}{\bar{f}_{ij,T}^{\sigma}}\right| = \left|\frac{f_{ij, T}^{\sigma} - \bar{f}_{ij, T}^{\sigma}}{\bar{f}_{ij, T}^{\sigma}}\right|~.
\end{equation}
Here, $f_{ij, T}^{\sigma}$ is the result from the numerical time-integration scheme, as described in Sec.~\ref{subsubsec:timeint}, while $\bar{f}_{ij, T}^{\sigma}$ denotes the reference value obtained from complex frequency-domain integration of the corresponding Green’s tensor component, cf. Appendix~\ref{app:RefValue}.

\begin{figure}[t]
	\includegraphics[width=\linewidth]{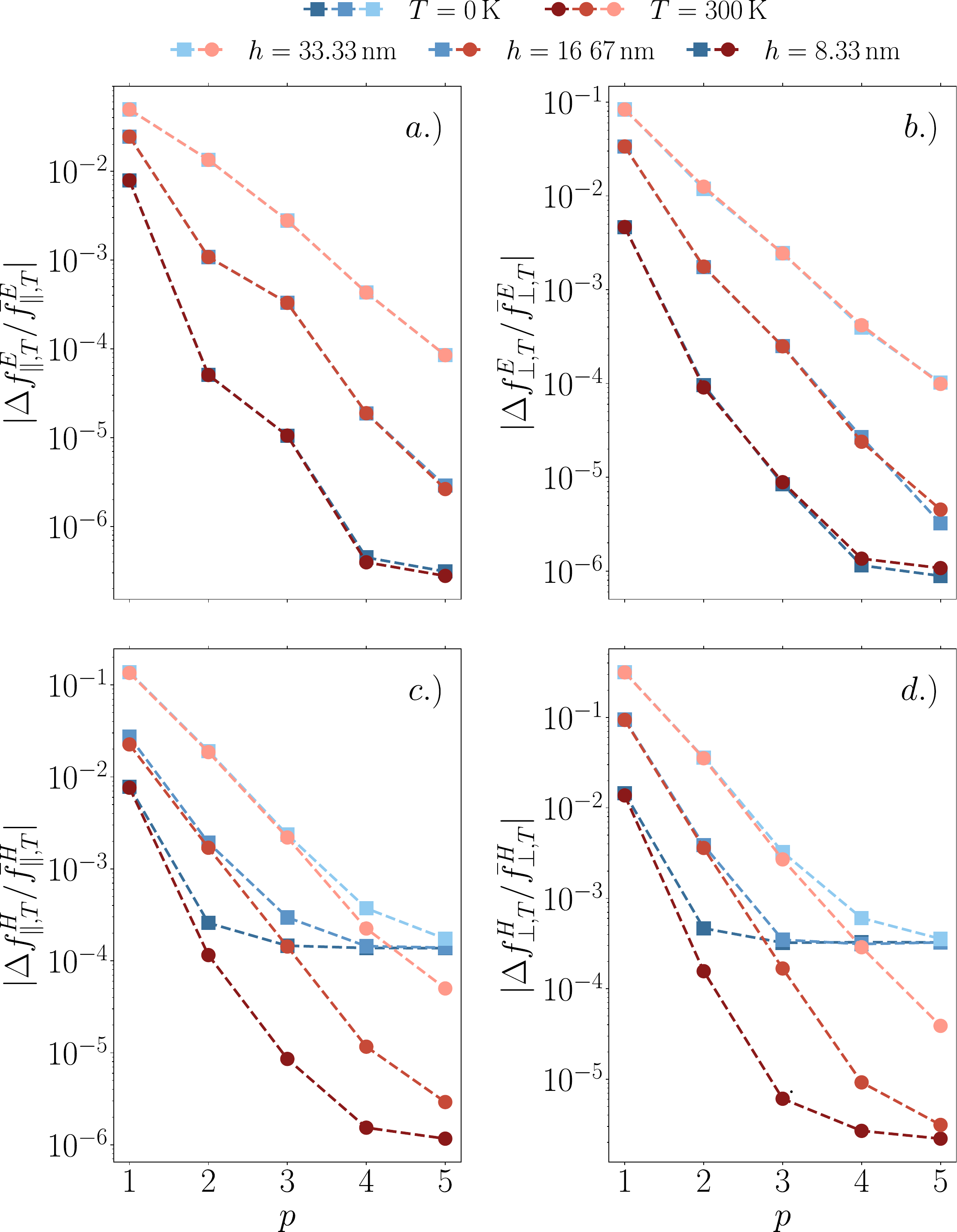}
	\caption{Relative error in the component-wise Casimir force per unit area at $L=100\,\textrm{nm}$, averaged over $22$ nominally identical simulations with total duration $\gamma t_\textrm{max}=30$, as a function of the polynomial order $p$ for different mesh resolutions. The mesh size $h$ is defined as the average edge length in the central region of the computational domain. Results are presented for both zero (blue) and room temperature (red). Individual contributions from the electromagnetic field components are shown separately: (a) and (b) correspond to the parallel ($\parallel$) and perpendicular ($\perp$) electric-field components, respectively, while (c) and (d) show the corresponding magnetic-field components.\label{fig:convstud_ms_po_T0}}
\end{figure}

The accuracy of the method is limited by different error sources, most notably the finite spatial discretization inherent to any mesh-based approach. To assess its effect, we perform a convergence study with respect to both mesh refinement, by uniformly reducing the average edge length $h$ through subdivision of each tetrahedral element, and polynomial refinement, by increasing the global polynomial order $p$ of the local basis functions~\cite{Kristensen2020}. The study is conducted for the same configuration as in the inset of Fig.~\ref{fig:cf_2hs_L_T}, with results averaged over $22$ nominally identical simulations obtained by displacing $\mathbf{r}_0$ within the plane $z=L/2$~\cite{Kristensen2023}. Although the physical configuration remains unchanged, shifting $\mathbf{r}_0$ modifies the distribution of the tetrahedral elements, leading to variations in the computed force. Averaging over these realizations reduces the sensitivity to a particular mesh placement and yields a more robust estimate of the spatial discretization error. Figure~\ref{fig:convstud_ms_po_T0} presents the averaged relative error as a function of the polynomial order for different mesh resolutions, where the average is taken in the logarithmic scale. Electric-field contributions in Figs.~\ref{fig:convstud_ms_po_T0}(a)~and~(b) exhibit the expected convergence behavior $\mathcal{O}(h^{p+1})$, consistent with theoretical predictions for conformal meshes, cf. Sec.~\ref{subsubsec:dgtd}. In contrast, the relative error for magnetic-field contributions at $T = 0\,\mathrm{K}$ in Figs.~\ref{fig:convstud_ms_po_T0}(c)~and~(d) plateaus beyond a certain polynomial order and mesh resolution, suggesting a transition to a regime dominated by other error sources rather than spatial discretization. Notice that this behavior is different at finite temperature, e.g., $T = 300\,\textrm{K}$, indicating that the faster decay of the temporal kernel relaxes this limitation. As discussed in Sec.~\ref{subsubsec:timeint}, time-domain integration introduces errors depending on the value of $t_{\textrm{max}}$, which we identify as the dominant remaining error source and examine in detail below.

\begin{figure}[t]
	\includegraphics[width=\linewidth]{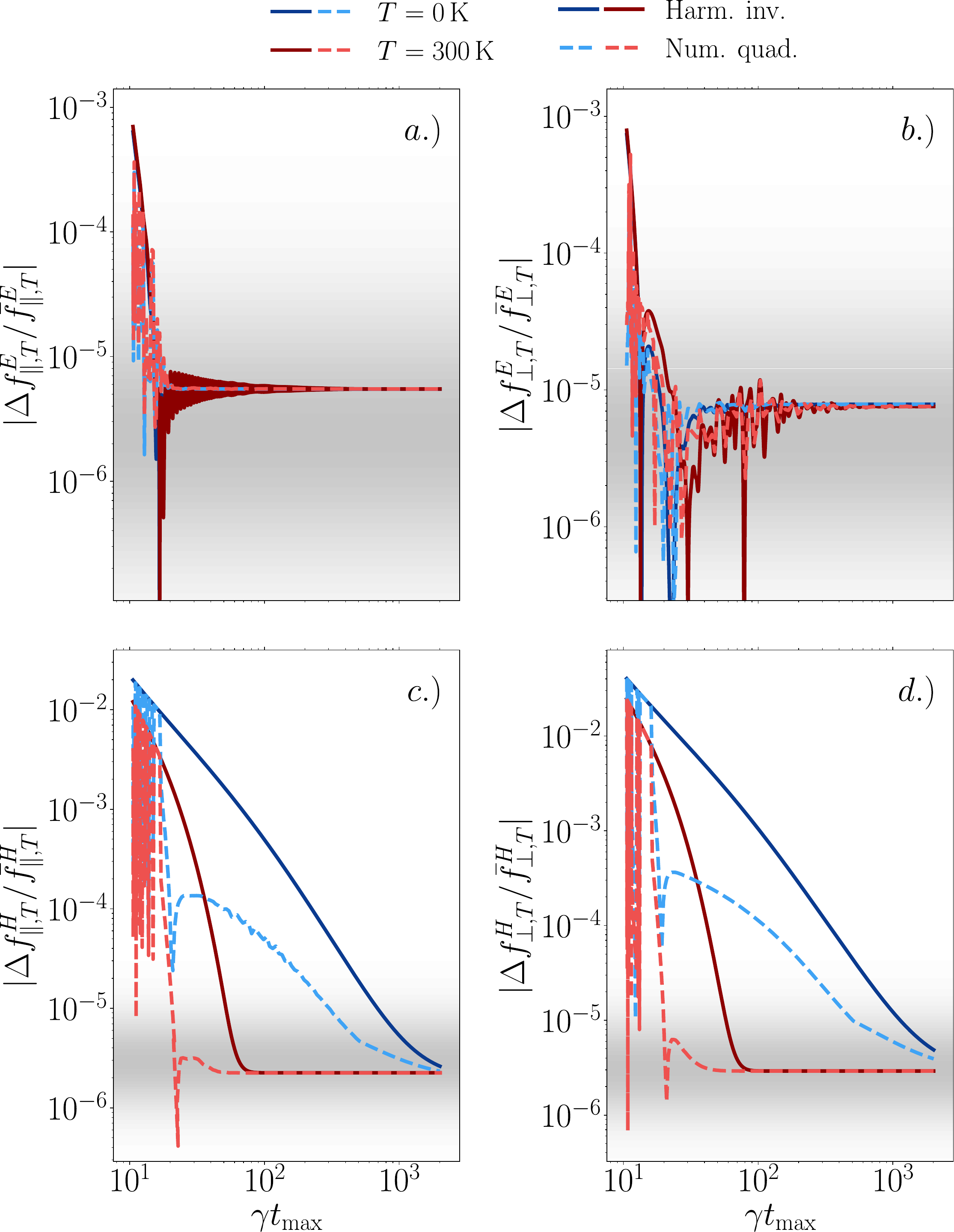}
	\caption{Relative error of the component-wise Casimir force per unit area as a function of the simulation time $t_\mathrm{max}$, for $L = 100\,\mathrm{nm}$, polynomial order $p = 5$, and mesh size $h = 16.67\,\mathrm{nm}$. Solid lines represent direct quadrature of the time-domain response over $[0,t_\mathrm{max}]$, while dashed lines include analytically integrated long-time contributions from $t_\mathrm{max}$ to $\tau\to\infty$, using the modal parameters obtained via harmonic inversion over $[t_0,t_\mathrm{max}]$, which are added to the quadrature result according to Eq.~\eqref{eq:f_time-HI}, effectively extending the integration to $t_\mathrm{max}\to\infty$. Results are shown for both zero (blue) and room temperature (red). The shaded region represents the deviation of the relative error from $22$ nominally identical simulations with computational domains where the roundtrip time exceeded the maximum $t_\mathrm{max}$, ensuring no spurious back-reflections within the simulation window. Panels (a) and (b) show the parallel ($\parallel$) and perpendicular ($\perp$) electric-field components, respectively, while panels (c) and (d) show the corresponding magnetic-field components.\label{fig:convstud_tmax}}
\end{figure}

To investigate this more precisely, we consider the impact of $t_{\textrm{max}}$ on the temporal integration given in Eq.~\eqref{eq:f_time-HI} for a fixed mesh size. This error, inherent to any time-domain method, can be reduced either by extending the simulation window, at the expense of increased computational effort, or by reconstructing the long-time behavior using parametric estimation methods such as the harmonic inversion. To quantify the effect of temporal truncation, Fig.~\ref{fig:convstud_tmax} shows the relative error as a function of $t_\textrm{max}$ at both zero and finite temperature, comparing direct quadrature integration over the interval $[0, t_\textrm{max}]$ with results obtained by adding harmonic-inversion-based long-time contributions according to Eq.~\eqref{eq:f_time-HI}. The latter reduces the computational effort required to achieve a certain level of accuracy. For the electric-field components shown in Figs.~\ref{fig:convstud_tmax}(a)~and~(b), increasing $t_\textrm{max}$ leads to rapid convergence toward a common residual error. This behavior suggests that, for sufficiently large $t_\mathrm{max}$, the dominant error shifts from temporal truncation to spatial discretization as addressed above. The electric-field components exhibit a weak dependence on long-time contributions, as the net dipole moment vanishes at large $\tau$, and the integrands in Eq.~\eqref{eq:f_time-HI} oscillate around the same asymptotic value, leading to faster convergence compared to the magnetic case. In contrast, the magnetic-field components are more sensitive to temporal truncation errors than their electric counterparts. In this case, relative errors are significantly reduced by the harmonic inversion approach, which captures the long-time behavior through the addition of the second integral of Eq.~\eqref{eq:f_time-HI}, compared to direct quadrature alone, considering only the first integral of Eq.~\eqref{eq:f_time-HI}, as shown in Figs.~\ref{fig:convstud_tmax}(c)~and~(d). Larger time windows improve the harmonic inversion accuracy, since they better resolve the slower temporal decay characteristic of the magnetic-field contributions~\cite{Johnson2004}, which also underlies their strong temperature dependence shown in Fig.~\ref{fig:convstud_ms_po_T0}. At $T = 0\,\textrm{K}$, the error takes longer to reach the residual value, while at $T = 300\,\textrm{K}$, the error rapidly saturates due to the faster decay of the kernel $\operatorname{Im}\left\{ g_{T}^{\sigma}(-\tau) \right\}$, cf. Appendix~\ref{app:asym} for details. As $t_\textrm{max}$ increases, the error in both direct quadrature and the harmonic inversion tends to the same residual error. It is important to note that the dips in $|\Delta f_{ij,T}^{\sigma}/\bar{f}_{ij,T}^{\sigma}|$ in Fig.~\ref{fig:convstud_tmax} correspond to sign changes in the error. Since these coincidental zero crossings are problem-dependent, no special significance should be attributed to the specific value of $\gamma t_\textrm{max}$ at which they occur.

As in Ref.~\cite{Kristensen2023}, we find that, in addition to temporal truncation and spatial discretization errors, the accuracy may be limited by reflections from the boundaries. To confirm that the residual error in Figs.~\ref{fig:convstud_ms_po_T0}~and~\ref{fig:convstud_tmax} is not dominated by spurious back-reflections from the finite termination of the computational domain, we perform control simulations using larger domains. In such cases, the radius $R$ is chosen such that the roundtrip time from the outer boundary to the evaluation point $\mathbf{r}_0$ exceeds the maximum simulation time $t_\textrm{max}$, preventing any boundary reflections from reaching $\mathbf{r}_0$ within the integration window. The results are averaged over $22$ nominally identical simulations with different tetrahedral distributions, and the standard deviation of the relative errors provides an independent estimate of the discretization error, represented by the shaded region in Fig.~\ref{fig:convstud_tmax}. This set of simulations provides a reference for assessing results obtained with reduced computational domains. Specifically, the residual error shown in Fig.~\ref{fig:convstud_tmax} falls within this spread, confirming that it is primarily governed by spatial discretization rather than by boundary reflections.

\subsection{Cylindrically symmetric object above a plane\label{subsec:cylSym}}

We now consider a system consisting of a finite, cylindrically symmetric object positioned above a half-space, separated by a surface-to-surface distance $L$. The coordinate system is chosen such that the $z$-axis coincides with the system's axis of symmetry and is oriented normal to the substrate, occupying $z \leq 0$. Owing to the cylindrical symmetry around the $z$-axis, the lateral components of the Casimir force cancel, resulting again in a net force directed along $\mathbf{e}_z$.

Once again, we choose a closed surface whose bottom face lies at $z = L/2$, midway between the object and the substrate, while the remaining faces extend to infinity, contributing negligibly due to the decay of the scattered fields, cf. Fig.~\ref{fig:surfint_hs-cyl}. The integral in Eq.~\eqref{eq:cf} therefore reduces to an effective open-surface integral over $S_{0}$ at $z = L/2$, with unit normal vector $\mathbf{n} = -\mathbf{e}_z$. Exploiting the system's axial symmetry, we work with cylindrical coordinates $\mathbf{r}\equiv(\rho, \varphi, z)$. The azimuthal dependence, $\varphi$, can be integrated analytically, reducing the surface integration of the stress tensor to a one-dimensional integral over the radial coordinate $\rho\in[0,\infty)$, which can be evaluated using a Gaussian quadrature scheme~\cite{Stroud1966}. Although Gauss--Laguerre quadrature can, in principle, handle semi-infinite intervals, in practice, it introduces nodes far from the object that contribute negligibly to the final value, while expanding the computational domain to impractical sizes. Truncating the integral to a finite interval and using a Gauss--Legendre quadrature scheme leads to an efficient computation~\cite{Trefethen2022}.

We leverage the adaptive-meshing capabilities of finite-element methods and sample the smoothly varying stress tensor at nonuniform radial positions corresponding to the nodes of a Gauss--Legendre quadrature of order $N$~\cite{Xiong2009,Dalvit2011}, bounded by a maximum cutoff radius $\rho_{\textrm{max}}$, beyond which the stress tensor is assumed to be negligible. This formulation enables efficient high-accuracy surface integration using standard Gaussian quadrature routines under a suitable change of variables while requiring only a small number of evaluation points. The resulting force, therefore, has an amplitude given by
\begin{equation}
	F_{z} \approx - \pi\rho_\textrm{max} \sum_{i=1}^{N} w_i \rho_i \langle \hat{\mathrm{T}}_{zz}(\rho_i, z=L/2, t) \rangle_T~,
\end{equation}
where $w_i$ and $x_i\in[-1,1]$ are the corresponding Gauss--Legendre weights and nodes, respectively, and $\rho_i=\rho_{\textrm{max}}[x_i+1]/2$ is the discretized radial coordinate.

\subsubsection{Surface integration\label{subsubsec:surfInt}}

As an illustrative example, we consider a finite metallic cylinder with radius $a$ and height $b = 2a$, positioned above a half-space. The material response is described by the Drude form given in Eq.~\eqref{eq:drude} for silver. The computational domain is shown in Fig.~\ref{fig:2hs_hs-cyl_mesh}(b), and the mesh is generated according to the procedure outlined in Sec.~\ref{subsubsec:timeint}. To resolve the cylinder’s curvature, the mesh size is set to $h = a/10$ near the lateral surface.

\begin{figure}[t]
	\includegraphics[width=\linewidth]{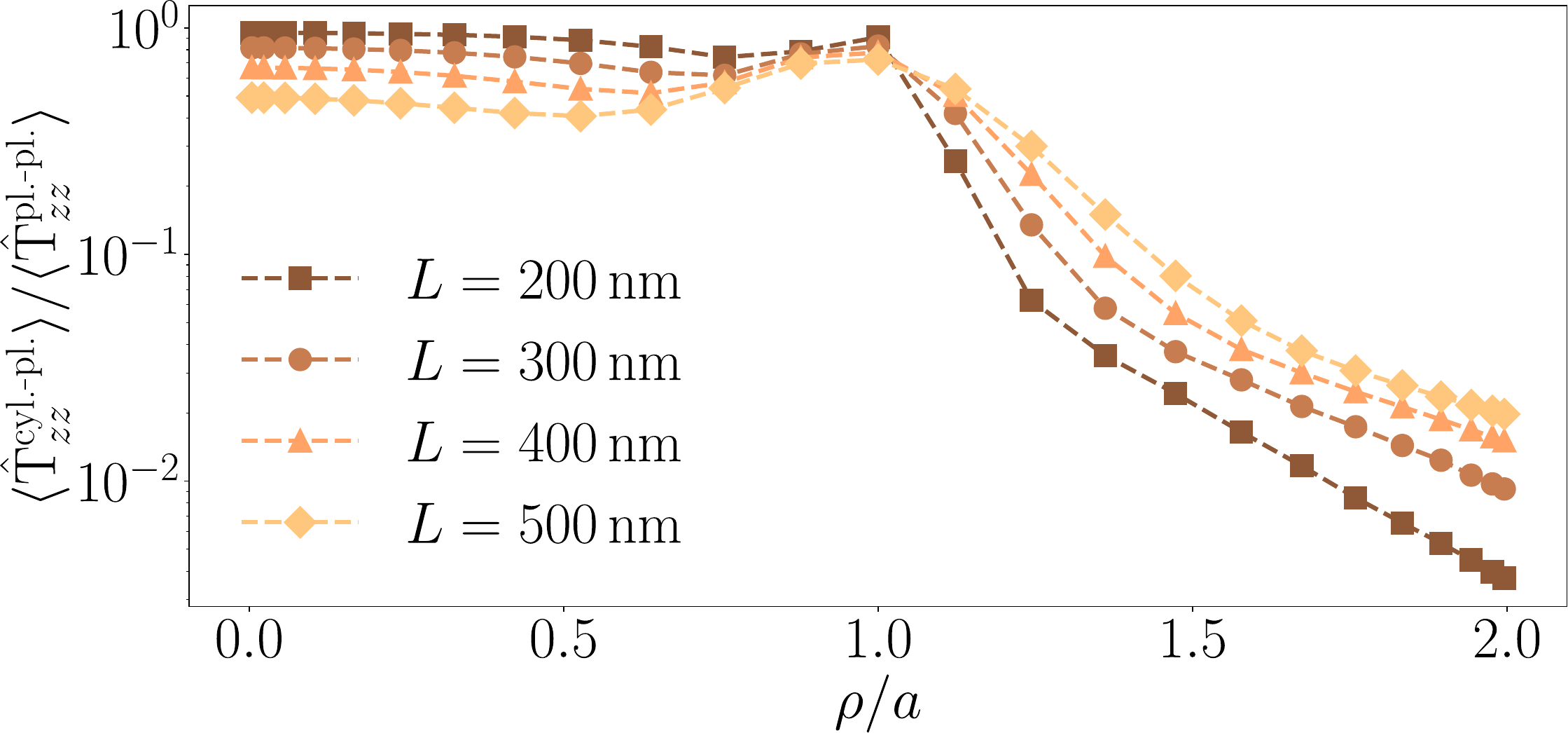}
	\caption{Spatial distribution of the Maxwell stress tensor for a cylinder of radius $a=500\,\textrm{nm}$ and height $b = 2a$ positioned at different separations above a half-space. Values are normalized to the plane--plane Maxwell stress tensor.\label{fig:surfint_hs-cyl}}
\end{figure}

The local Casimir force contributions per unit area, $f_{ij, T}^\sigma(\rho, z=L/2)$, are sampled at radial positions determined by a $16$-point Gauss--Legendre quadrature within the domain $\rho \leq \rho_\textrm{max}=a + L$. At each quadrature node, a point dipole is placed, and the corresponding scattered fields are evolved in time. The field amplitudes are recorded at the dipole positions and integrated over time using the scheme described in Sec.~\ref{subsubsec:timeint}. The total Maxwell stress tensor, shown in Fig.~\ref{fig:surfint_hs-cyl}, is obtained by summing the local Casimir force contributions per unit area, as given by Eq.~\eqref{eq:strTen_time}. We note that near the cylinder center, at $\rho \ll a$, the configuration resembles the translationally invariant plane--plane geometry discussed in the previous section, and thus the stress tensor remains nearly uniform. For $\rho\sim a$, a peak appears, which can be attributed to the increasing relevance of the scattered fields coming from the lateral surface of the cylinder. For $\rho\gg a$, the stress tensor decays exponentially at a rate scaling as the inverse of $L$, thereby justifying a posteriori the choice of $\rho_\textrm{max}$. The spatial integration of these contributions yields the resulting Casimir force. Given the rapid convergence of Gaussian quadrature techniques~\cite{Stroud1966}, and for sufficiently large $\rho_{\textrm{max}}$, we expect the dominant numerical error to arise from the spatial discretization of the mesh. For non-conformal meshes, such as in the present case, the convergence rate scales as $\mathcal{O}(h^2)$~\cite{Viquerat2015}.

\begin{figure}[b]
	\includegraphics[width=\linewidth]{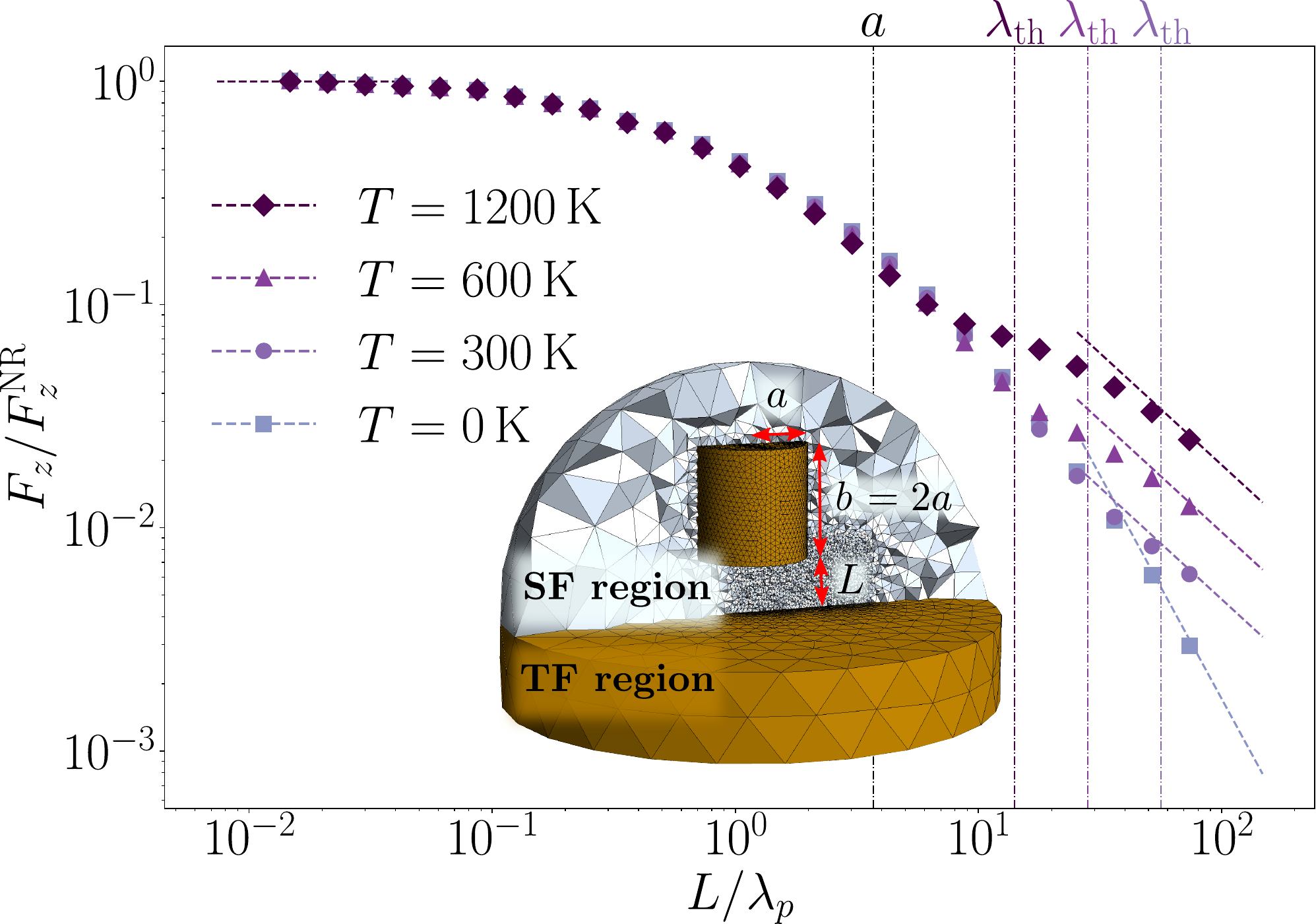}
	\caption{Casimir force between a cylindrical object and a half-space as a function of their separation $L$. The cylinder has radius $a = 500\,\textrm{nm}$ and height $b = 2a$. The force is normalized to the nonretarded limit of the Casimir pressure in Eq.~\eqref{eq:asympLifshitzFormula}, multiplied by the area of the cylinder's base. Symbols denote numerical results, while dashed lines represent semi-analytical asymptotic predictions. At short separations, the numerical results recover the $\sim -1 / L^3$ behavior predicted by Eq.~\eqref{eq:asympLifshitzFormula} and are independent of temperature, whereas at large separations they reproduce the $\sim -1 / L^5$ and $\sim -1 / L^4$ dependencies expected from the asymptotic expressions for the Casimir--Polder interaction in Eqs.~\eqref{eq:CasimirPolderRetardedT0} and \eqref{eq:CasimirPolderRetardedT} at zero and finite temperature, respectively. The inset shows a representative mesh of the cylinder--plane geometry generated with Gmsh~\cite{Geuzaine2009}.\label{fig:cf_hs-cyl_L_a_T}}
\end{figure}

Figure~\ref{fig:cf_hs-cyl_L_a_T} shows the force as a function of the separation $L$, normalized by the pressure given in Eq.~\eqref{eq:asympLifshitzFormula} multiplied by the area of the cylinder's base, $A = \pi a^{2}$. This value corresponds to the nonretarded limit of the proximity-force approximation (PFA), which is obtained from the Lifshitz formula in Eq.~\eqref{eq:LifshitzFormula} multiplied by $A$. Numerical results (symbols) are compared with asymptotic predictions (dashed lines) in Fig.~\ref{fig:cf_hs-cyl_L_a_T}, where different temperatures are considered for a cylindrical object with a fixed radius of $a = 500\,\textrm{nm}$.

Although for this configuration no analytical solution exists, the limiting behaviors for both short- and long-distance regimes can be reliably predicted based on physical considerations. At separations $L \ll a$, when the distance is much smaller than the radius of the cylindrical structure, the interaction should approach that of two half-spaces. This behavior is consistent with that predicted by the PFA~\cite{Fosco2012,Bimonte2012}. For our configuration, this means that the impact of the scattering arising from the cylinder's boundaries can be neglected, leading to a force given by the Lifshitz pressure on the effective area of the system, i.e., the area of the cylinder's base $A$. Consequently, when $L\ll \lambda_p$ and $L\ll a$, the numerically evaluated Casimir force becomes temperature independent and scales as $F_z \sim - 1/L^{3}$, thereby justifying the choice of normalization used in Fig.~\ref{fig:cf_hs-cyl_L_a_T}.

\begin{figure}[t]
	\includegraphics[width=\linewidth]{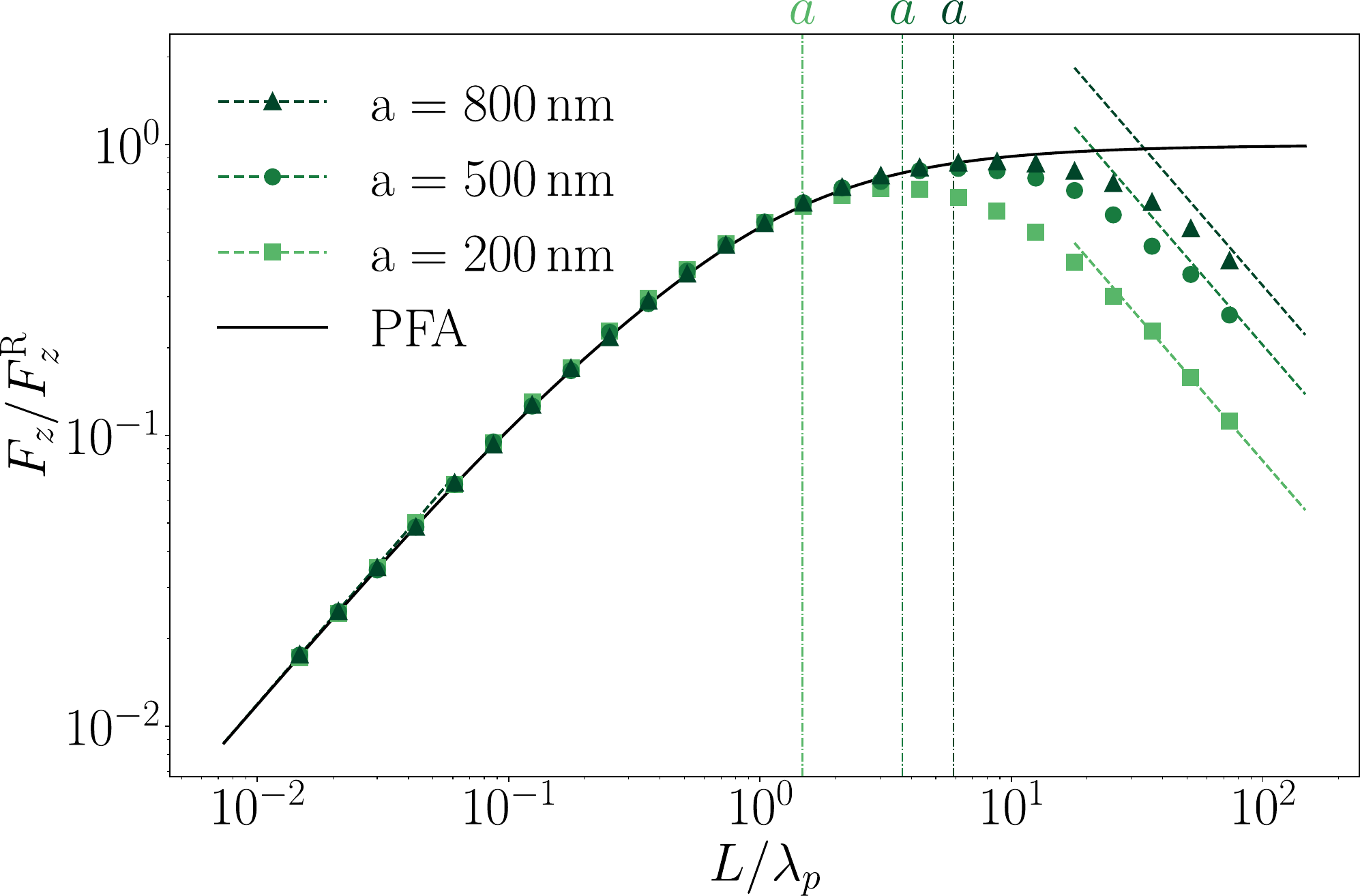}
	\caption{Casimir force between a cylindrical object and a half-space as a function of their separation $L$. The temperature is fixed at $T = 0\,\textrm{K}$, while different cylinder radii $a$, and corresponding heights $b=2a$, are considered. The force is normalized to the retarded limit of the Casimir pressure in Eq.~\eqref{eq:CasimirP}, multiplied by the area of the cylinder's base. As in Fig.~\ref{fig:cf_hs-cyl_L_a_T}, symbols denote numerical results, dashed lines represent asymptotic predictions, and solid lines semi-analytical approximations based on the PFA. At short separations, the numerical results recover the $\sim -1 / L^3$ behavior predicted by Eq.~\eqref{eq:asympLifshitzFormula}, whereas at large separations they reproduce the $\sim -1 / L^5$ dependence from the Casimir--Polder interaction in Eq.~\eqref{eq:CasimirPolderRetardedT0}, with $\alpha(0)$ varying according to $a$.\label{fig:cf_hs-cyl_L_a_T0}}
\end{figure}

When the distance becomes larger than the radius of the compact object, the accuracy of the PFA degrades, losing its validity, and the numerics deviate from the semi-analytical estimate, cf. Fig.~\ref{fig:cf_hs-cyl_L_a_T0}. In this regime, the interaction is increasingly governed by the cylinder’s multipole scattering contributions~\cite{Bohren2008}. For $L\gg a$, when the cylinder dimensions are small compared to its distance from the surface, higher-order multipoles can be neglected, and the dipolar response dominates the scattered electromagnetic radiation. In this regime, the system can be effectively described in terms of a microscopic object interacting with a surface, and the Casimir force gradually transitions into the Casimir--Polder force~\cite{Casimir1948}. Unlike the Lifshitz expression in Eq.~\eqref{eq:LifshitzFormula}, the Casimir--Polder interaction is directly connected to the polarizability of the microscopic object. While no closed-form expression exists for the polarizability of a small cylinder~\cite{Venermo2005}, a reliable approximation is provided by relating it to that of a volume-equivalent spheroid~\cite{Rodriguez2013}. Within this approach and for this specific case, a cylindrical object of equal height and diameter can be roughly approximated by a sphere of volume $V = \pi a^2 b$. Under these conditions, the frequency-dependent polarizability follows the Clausius--Mossotti relation~\cite{Jackson1975,Intravaia2011}
\begin{equation}
	\alpha(\omega) \approx 3\epsilon_{0}V \frac{\epsilon(\omega) - 1}{\epsilon(\omega) + 2}~,
\end{equation}
where $\epsilon(\omega)$ is the dielectric function given in Eq.~\eqref{eq:drude}, and thus the polarizability features a resonance at $\omega_{r}\approx \omega_{p}/\sqrt{3}$, for $\Gamma\ll \omega_{p}$.

The preceding considerations enable us to estimate the behavior of the force at large separations. At finite temperature, in the thermal regime, i.e., $L \gg \lambda_{\rm th}$, the Casimir--Polder interaction is dominated as above by the thermal fluctuations, yielding the thermal limit~\cite{Intravaia2011}
\begin{equation}
\label{eq:CasimirPolderRetardedT}
F_z \sim - \frac{3}{16\pi}\frac{k_\textrm{B}T}{\epsilon_{0}}\frac{\alpha(0)}{L^{4}}~.
\end{equation}
The change in the dimensionality of the object and, in particular, the restriction of its lateral extent give rise to a power law, which decays one order faster than the Casimir force in the corresponding limit. This is visible in the trend of the numerical data in Fig.~\ref{fig:cf_hs-cyl_L_a_T}, which bend downwards compared to what is observed in Fig.~\ref{fig:cf_2hs_L_T}. With increasing separation, the data converge to the asymptotic limit in Eq.~\eqref{eq:CasimirPolderRetardedT}, which is set by the system’s temperature, validating both the numerics and the underlying physical reasoning. At zero temperature, the interaction features a different asymptotic behavior at large separations. The retarded regime of the Casimir--Polder force is defined by $L\gg \lambda_{r}=2\pi \mathrm{c}/\omega_{r}\approx\sqrt{3}\lambda_{p}$, approximately overlapping with that of the Casimir force. Therefore, for $L \gg \lambda_p$ and $L \gg a$, the interaction behaves according to~\cite{Buhmann2007,Intravaia2011,Buhmann2012}
\begin{equation}
\label{eq:CasimirPolderRetardedT0}
F_z \sim -\frac{3}{8\pi^{2}}\frac{\hbar \mathrm{c}}{\epsilon_{0}}\frac{\alpha(0)}{L^{5}}~,
\end{equation}
which again features a characteristic power law, decaying one order faster than the Casimir force at zero temperature in the corresponding limit, cf. Eq.~\eqref{eq:CasimirP}. We observe this behavior in Fig.~\ref{fig:cf_hs-cyl_L_a_T0}, where we considered cylinders with different values of $a$ at $T=0\,\textrm{K}$. To highlight the deviation from the PFA prediction (solid line), the normalization is chosen with respect to the Casimir pressure in Eq.~\eqref{eq:CasimirP} multiplied by the area of the cylinder's base. For each cylinder size, the numerical data (symbols) progressively approach the asymptotic behavior described by Eq.~\eqref{eq:CasimirPolderRetardedT0}. As expected, the smaller the value of $a$, the shorter the distance at which the numerics start to deviate from the PFA. The dependence of the asymptotic expression on $a$ follows directly from the relation $\alpha(0)=6\epsilon_{0}\pi a^{3}$.

Finally, it is worth stressing that, although the asymptotic behaviors can be predicted and analytically evaluated based on physical grounds, the intermediate regime lacks closed-form expressions and precise estimates. In this range, which is relevant for experimental setups where the interplay between geometry, dispersion, and temperature determines the force value, sufficiently accurate predictions can only be achieved through numerical methods like the one presented in this work.

\section{Conclusions\label{sec:conclusions}}

In this work, we have presented a high-accuracy, time-domain finite-element-based framework for computing Casimir forces within the Maxwell stress tensor formalism. The numerical scheme employs the discontinuous Galerkin time-domain (DGTD) method to evaluate Casimir interactions between arbitrarily shaped objects described by general material models, at finite temperature. By expressing the electromagnetic Green’s tensor in terms of the system's response to dipolar excitations, the stress tensor is recast into a set of classical scattering problems driven by electric and magnetic dipole sources in the time domain. Benchmark calculations for the Casimir interaction between two half-spaces have shown excellent agreement with the Lifshitz formula across multiple length scales. Relevant steps in achieving this outcome were the design of a specific temporal profile for the electric and magnetic sources and the development of a harmonic-inversion-based procedure to extrapolate the long-time behavior of the electromagnetic fields. Regarding the convergence of the scheme, our analysis revealed that it is primarily limited by temporal truncation and spatial discretization.

We further applied the framework to the Casimir interaction between a finite metallic cylindrical object and a substrate, a geometry lacking an analytical solution. The numerical results reproduced the expected asymptotic behavior in both short- and long-distance limits, while providing quantitative predictions in the intermediate regime where geometry, dispersion, and temperature effects interact in nontrivial ways. One of the primary technical challenges of our approach was the evaluation of the Maxwell stress tensor flux over a closed integration surface. For the specific configuration in this work, we used a Gaussian quadrature scheme to reach the desired level of accuracy within a relatively small number of quadrature points. While the present calculation benefits from cylindrical symmetry, the framework can also be extended to less symmetric configurations using optimized two-dimensional Gaussian quadrature or cubature schemes~\cite{Cools1993,Cools1999,Cools2003}. Although such an extension would increase the number of required simulations, the independence of each run makes the scheme ideally suited for parallelization on high-performance computing clusters. This highlights the flexibility and accuracy of the DGTD approach for configurations that lie beyond the scope of commonly used approximations, such as the proximity-force approximation, and more rigorous scattering multipole expansions.

In summary, the DGTD method provides an efficient and accurate alternative for computing Casimir-type forces in the time domain. Its ability to accommodate arbitrary geometries, material models, and finite temperature makes it particularly well suited for studying fluctuation-induced forces in micro- and nanoscale systems. Future developments may include the implementation of curvilinear elements to preserve high-order convergence~\cite{Viquerat2015}, which would be especially beneficial for spherical geometries commonly encountered in experimental setups. On the material side, the approach can be extended to incorporate spatially dispersive models in nanostructured geometries, which are becoming increasingly relevant in Casimir physics~\cite{EsquivelSirvent2006,Intravaia2013}. The DGTD method naturally accommodates spatial dispersion through ADEs~\cite{Wegner2023,Kristensen2023}, offering important advantages over boundary element methods, which are limited by their surface-only treatment, and over finite-difference time-domain methods, which struggle to resolve subnanometer variations on a fixed grid with multiple length scales.

\begin{acknowledgments}
C.M.F. and K.B. acknowledge support from the German Research Foundation (DFG) within the framework of the Collaborative Research Center (CRC) 1375, Project A06 (Project ID 398816777). F.I. acknowledges BERLIN QUANTUM, an initiative endowed by the Innovation Promotion Fund of the city of Berlin, for financial support.
\end{acknowledgments}

\appendix

\section{Asymptotic behavior of $\operatorname{Im}\left\{g_{T}^{\sigma}(-\tau)\right\}$\label{app:asym}}

The kernel $\operatorname{Im}\left\{g_{T}^{\sigma}(-\tau)\right\}$ in Eq.~\eqref{eq:img_time} is a central quantity of the approach. In this appendix, we examine its asymptotic behavior in the short-time, long-time, as well as finite- and zero-temperature regimes, for electric ($\sigma = E$) and magnetic ($\sigma = H$) dipolar sources.

In the short-time limit, i.e., for $\tau \ll 1/\gamma$ and $\tau \ll1/\omega_{\rm th}$, with $\omega_{\rm th} = 2\pi k_\textrm{B} T/\hbar$, the polylogarithm can be asymptotically expanded as $\operatorname{Li}_{-l}(\mathrm{e}^{-\omega_{\rm th} \tau}) \sim l! / (\omega_{\rm th} \tau)^{l+1}$~\cite{Wood1992}. In this regime, the leading-order contribution to the expression in Eq.~\eqref{eq:img_time} is provided by the lowest-order term, i.e., $l = s+1$ for $\sigma=E$ and $l = s$ for $\sigma=H$. Inserting it in Eq.~\eqref{eq:img_time} then yields the short-time behavior
\begin{equation}
	\lim\limits_{\tau\to0}\operatorname{Im}\left\{g_{T}^{\sigma}(-\tau)\right\} \sim \tilde{g}_{0}^\sigma\frac{(s+1)!}{(\gamma \tau)^{s+2}}~,
	\label{eq:img_tauto0}
\end{equation}
with $\tilde{g}_{0}^\sigma = \gamma^3 / 2\pi j_0^\sigma s!$. Note that Eq.~\eqref{eq:img_tauto0} does not depend on the temperature and is the same for both the magnetic and electric sources, revealing a source-type-independent power-law divergence at short times~\cite{Kristensen2023}. For our choice $s=4$, this coincides with the $1/\tau^{6}$ behavior visible in the insets of panels (c) in Fig.~\ref{fig:timeint_2hs_f}.

In the opposite limit, i.e., for $\tau \gg 1/\gamma$ and $\tau \gg 1/\omega_{\rm th}$, the condition $\mathrm{e}^{-\omega_{\rm th} \tau} \ll 1$ holds, leading to an exponential decay of the polylogarithm function. As a result, the Matsubara series in Eq.~\eqref{eq:img_currDens} is dominated by its first term, which leads to the asymptotic behavior
\begin{equation}
	\begin{aligned}
		\lim\limits_{\tau\to\infty}\operatorname{Im}\{g_{T}^{\sigma}(-\tau)\}& \sim
		\\
		\tilde{g}_{0}^\sigma
		&\begin{cases}
			\displaystyle\frac{\omega_{\rm th}}{2\gamma} & \sigma = E~,  \\[12pt]
			\displaystyle \mathrm{e}^{-\omega_{\rm th} \tau} \sum_{l=0}^{s} \binom{s}{l}
			\left( \frac{\omega_{\rm th}}{\gamma} \right)^{l+2} & \sigma = H~,
		\end{cases}
	\end{aligned}
	\label{eq:img_tautoinf}
\end{equation}
so that the electric kernel approaches a constant, whereas the magnetic kernel decays exponentially.

As the temperature approaches zero, the spacing between adjacent Matsubara frequencies becomes vanishingly small, and the discrete sum in Eq.~\eqref{eq:img_currDens} approaches an integral,
\begin{equation}
	\omega_{\rm th}\sideset{}{'}\sum_{m=0}^{\infty} \longrightarrow \int\limits_{0}^{\infty} \mathrm{d}\xi~.
\end{equation}
In this regime, the kernel in Eq.~\eqref{eq:img_time} reduces to a sum of pure power-law terms, expressed as
\begin{equation}
	\begin{aligned}
		\lim\limits_{T\to0}\operatorname{Im}\left\{g_{T}^{\sigma}(-\tau)\right\} = & \\
		\tilde{g}_{0}^\sigma&\begin{cases}
			\displaystyle\sum_{l=0}^{s+1} \binom{s+1}{l} \frac{l!}{(\gamma \tau)^{l+1}} & \sigma = E~, \\[12pt]
			\displaystyle\sum_{l=0}^{s} \binom{s}{l} \frac{(l+1)!}{(\gamma \tau)^{l+2}} & \sigma = H~.
		\end{cases}
	\end{aligned}
\end{equation}
At short times, $\tau\ll1/\gamma$, these reproduce the divergent behavior described in Eq.~\eqref{eq:img_tauto0}. In the long-time limit, $\tau \gg 1/\gamma$, only the lowest-order terms contribute, yielding
\begin{equation}
	\lim_{\substack{T \to 0 \\ \tau \to \infty}}\operatorname{Im}\left\{g_{T}^{\sigma}(-\tau)\right\} \sim \tilde{g}_{0}^\sigma
	\begin{cases}
		\displaystyle \frac{1}{\gamma \tau} & \sigma = E~, \\[12pt]
		\displaystyle \frac{1}{(\gamma \tau)^2} & \sigma = H~,
	\end{cases}
	\label{eq:LongTimesT0}
\end{equation}
revealing a power-law decay, in contrast to the constant or decaying exponential behavior observed at finite temperature. As above, the long-time asymptotic trends are confirmed by the numerical evaluation in the inset of panels (c) of Fig.~\ref{fig:timeint_2hs_f}.

\section{The role of $\gamma$ and $\chi^\sigma$\label{app:sourceParam}}

The time-dependent current density in Eq.~\eqref{eq:currDens_time} depends on two pulse-shaping parameters: $\gamma$, which determines the temporal and spectral characteristics of the pulse, and $\chi^\sigma$, a discrete source-specific parameter that modulates the temporal profile of the pulse and influences the decay behavior of the integration kernel, as discussed in Appendix~\ref{app:asym}. Although both parameters are formally arbitrary within the theoretical description, a comprehensive understanding of their influence is relevant for ensuring numerical convergence, computational efficiency, and accuracy.

In the time domain in Eq.~\eqref{eq:currDens_time}, $\gamma$ acts as an exponential envelope controlling the pulse width, while in the frequency domain in Eq.~\eqref{eq:currDens_freq}, it determines the spectral bandwidth. Increasing $\gamma$ produces temporally narrower but spectrally broader pulses. A priori, a large value of $\gamma$ is desirable for broadband excitations~\cite{Rodriguez2009}. However, in addition to shaping the temporal and spectral properties of the pulse, $\gamma$ also influences the convergence behavior of the numerical scheme through the kernel $\operatorname{Im}\{g_{T}^{\sigma}(-\tau)\}$. As discussed in Appendix~\ref{app:asym}, this diverges as $\sim1/\tau^{s+2}$ for $\tau\ll 1/\gamma$ and decays as $\sim1/\tau^{2-\chi^\sigma}$ for $1/\gamma \ll \tau \ll 1/\omega_{\rm th}$, where $\chi^\sigma$ can be either $0$ or $1$. These regimes intersect at a characteristic timescale $t^\star = [(s+1)!]^{1/(s+\chi^\sigma)} / \gamma$, which delineates the transition from the rising flank of the integrand, $I^{\sigma}_{ij, T}$, at short times ($\tau<t^\star$) to the trailing edge at long times ($\tau>t^\star$), cf. panels (d) of Fig.~\ref{fig:timeint_2hs_f}. To prevent premature suppression of relevant field contributions, $t^\star$ is set to align with the physical arrival time $L / \mathrm{c}$. Consequently, the optimal choice of $\gamma$ must balance a broadband excitation with a proper alignment of the kernel and the scattered-field arrival time. In practice, we find that setting $\gamma \lesssim \mathrm{c}/2L$ has proven effective across a wide range of length scales.

At finite temperature, the timescale $1/\omega_{\rm th}$ modifies the long-time behavior of the kernel. For $\tau \gg 1/\omega_{\rm th}$, thermal effects dominate, and, depending on $\chi^\sigma$, the kernel either approaches a constant or decays exponentially, as seen in Eq.~\eqref{eq:img_tautoinf}. For electric sources with $\chi^E = 1$~\cite{Kristensen2023}, the kernel becomes time-independent in the long-time limit. The convergence of the time integration in Eq.~\eqref{eq:f_time} in this case is primarily controlled by the decay behavior of the scattered electric field. However, since the choice $\chi^E = 1$ implies that the integral over positive times of the current profile in Eq.~\eqref{eq:currDens_time} vanishes as a result of the lack of net dipole moment at long times, the time integral in Eq.~\eqref{eq:f_time} oscillates around its asymptotic value, becoming less sensitive to the truncation time, as seen in the inset of Fig.~\ref{fig:timeint_2hs_f}(e), left column. 

A different situation arises for magnetic sources. If $\chi^H=1$ is retained, the kernel again approaches a constant at long times $\tau \gg 1/\omega_{\textrm{th}}$. In combination with the slow decay of the scattered magnetic field, due to the slowly decaying nonoscillatory modes in the system~\cite{Intravaia2009,Intravaia2010,Henkel2010}, as seen in the right column of Fig.~\ref{fig:timeint_2hs_f}(b), this compromises the numerical convergence. To improve performance in the time domain at nonzero temperature, we therefore set $\chi^H = 0$. This forces the term $\xi_m / j^\sigma(\mathrm{i}\xi_m)$ in Eq.~\eqref{eq:img_currDens} to vanish at $\xi_m = 0$, resulting in an exponentially decaying kernel at long times, cf. Eq.~\eqref{eq:img_tautoinf}.

\section{Benchmark quantities\label{app:RefValue}}

The reference quantities $\bar{f}^{\sigma}_{ij, T}$ are obtained through a complex-frequency-domain evaluation of the corresponding Green's tensor components and are employed to assess the accuracy of the numerical time-integration scheme. In analogy with Eqs.~\eqref{eq:strTen_freq} and \eqref{eq:typeFields_freq}, the Maxwell stress tensor can be written in terms of the electromagnetic Green's tensor in the frequency domain as
\begin{equation}
	\langle \hat{\mathrm{T}}_{ij}(\mathbf{r}, t) \rangle_T = -2k_\textrm{B}T\sum_{\sigma}\sideset{}{'}\sum_{m=0}^{\infty}\frac{\xi_{m}^2}{\mathrm{c}^2}\mathcal{G}^{\sigma}_{ij}(\mathbf{r}, \mathrm{i}\xi_{m})~,
\end{equation}
where
\begin{equation}
	\mathcal{G}^{\sigma}_{ij}(\mathbf{r}, \mathrm{i}\xi_{m}) = G_{ij}^{\sigma}(\mathbf{r},\mathrm{i}\xi_{m})- \frac{\delta_{ij}}{2}\sum_{k}G_{kk}^{\sigma}(\mathbf{r},\mathrm{i}\xi_{m})~.
\end{equation}
The summation over imaginary frequencies results from the Wick rotation of the real-frequency stress tensor. This converts the oscillatory integrand into an exponentially decaying one, which can be easily handled numerically~\cite{Ford1993}. The component-wise contributions to the local Casimir force per unit area are therefore
\begin{equation}
	\bar{f}^{\sigma}_{ij, T}(\mathbf{r}) = -2k_\textrm{B}T\sideset{}{'}\sum_{m=0}^{\infty}\frac{\xi_{m}^2}{\mathrm{c}^2}G^{\sigma}_{ij}(\mathbf{r}, \mathrm{i}\xi_{m})~.
	\label{eq:fbar}
\end{equation}
However, analytical expressions for the Green's tensor $\underline{G}^{\sigma}$ exist only for a few highly symmetric geometries. One such case is the system of two parallel half-spaces discussed in Sec.~\ref{subsec:transSym}. Due to the translational symmetry, the Green's tensor depends only on the perpendicular coordinate $z$, and azimuthal symmetry ensures that the only nonvanishing components are $G_{xx}^{\sigma}=G_{yy}^{\sigma}\equiv G_{\parallel}^{\sigma}$ and $G_{zz}^{\sigma}\equiv G_{\perp}^{\sigma}$. For two identical half-spaces separated by a vacuum gap of width $L$, the electric Green's tensor reads~\cite{Tomas1995}
\begin{widetext}
	\begin{subequations}
		\begin{eqnarray}
			G_\parallel^E(z,\mathrm{i}\xi_m)=\frac{1}{8\pi}\int\limits_{0}^{\infty}\mathrm{d}q\,\frac{q}{\kappa}\Bigg[\left(\frac{\kappa \mathrm{c}}{\xi_m}\right)^2&&\frac{2r_{\mathrm{TM}}^2(\mathrm{i}\xi_{m},q)\mathrm{e}^{-2\kappa L}-r_{\mathrm{TM}}(\mathrm{i}\xi_{m},q)\mathrm{e}^{-2\kappa z}-r_{\mathrm{TM}}(\mathrm{i}\xi_{m},q)\mathrm{e}^{-2\kappa(L-z)}}{1-r_{\mathrm{TM}}^2(\mathrm{i}\xi_{m},q)\mathrm{e}^{-2\kappa L}}\\+&&\frac{2r_{\mathrm{TE}}^2(\mathrm{i}\xi_{m},q)\mathrm{e}^{-2\kappa L}+r_{\mathrm{TE}}(\mathrm{i}\xi_{m},q)\mathrm{e}^{-2\kappa z}+r_{\mathrm{TE}}(\mathrm{i}\xi_{m},q)\mathrm{e}^{-2\kappa(L-z)}}{1-r_{\mathrm{TE}}^2(\mathrm{i}\xi_{m},q)\mathrm{e}^{-2\kappa L}}\Bigg]\nonumber~,\label{eq:gt_2hs_par}\\[12pt]
			G_\perp^E(z,\mathrm{i}\xi_m)=-\frac{1}{4\pi}\int\limits_{0}^{\infty}\mathrm{d}q\,\frac{q}{\kappa}\left(\frac{q \mathrm{c}}{\xi_m}\right)^2&&\frac{2r_{\mathrm{TM}}^2(\mathrm{i}\xi_{m},q)\mathrm{e}^{-2\kappa L}+r_{\mathrm{TM}}(\mathrm{i}\xi_{m},q)\mathrm{e}^{-2\kappa z}+r_{\mathrm{TM}}(\mathrm{i}\xi_{m},q)\mathrm{e}^{-2\kappa(L-z)}}{1-r_{\mathrm{TM}}^2(\mathrm{i}\xi_{m},q)\mathrm{e}^{-2\kappa L}}~,\label{eq:gt_2hs_perp}
		\end{eqnarray}
	\end{subequations}
	\label{eq:gt_2hs}
\end{widetext}
for $0\le z \le L$. Here, $q=|\mathbf{q}|$ is the in-plane wave-vector and $\kappa=\sqrt{q^{2}+\xi_{m}^{2}/\mathrm{c}^{2}}$ is the perpendicular component in vacuum. For planar geometries and due to the symmetry of Maxwell's equations, the magnetic Green's tensor is obtained from the electric one by multiplying it by $\mathrm{c}^2$ and exchanging the $\textrm{TE}$ and $\textrm{TM}$ reflection coefficients ($r_{\mathrm{TE}}\leftrightarrow r_{\mathrm{TM}}$) in Eqs.~\eqref{eq:gt_2hs}~\cite{Dalvit2011}. In practice, the reference values $\bar{f}^{\sigma}_{ij, T}$, used in Sec.~\ref{subsubsec:convstud}, are computed by numerically evaluating the Matsubara sum in Eq.~\eqref{eq:fbar} with Eqs.~\eqref{eq:gt_2hs}. The resulting values provide a reliable benchmark for comparison with the time-domain simulations.

%

\end{document}